%% file: 0_Instrument-master.tex
\documentclass[a4paper,11pt]{article}
\usepackage{jcappub} 

\usepackage{lineno}
\usepackage{xcolor}
\usepackage{units}
\usepackage{url}
\usepackage{caption}
\usepackage{subfigure} 
\usepackage{amsmath}
\usepackage{gensymb}
\usepackage{textcomp}

\title{The Extreme Universe Observatory on a Super-Pressure Balloon II: Mission, Payload, and Flight}

  \author[ld]{J.H.~Adams~Jr.,}
  \author[cb]{D.~Allard,}
  \author[ld]{P.~Alldredge,}
  \author[le]{L.~Anchordoqui,}
  \author[ed,eh]{A.~Anzalone,}
  \author[lh]{M.~Bagheri,}
  \author[ek,cb]{M.~Battisti,}
  \author[ea,eb]{R.~Bellotti,}
  \author[ib]{A.A.~Belov,}
  \author[ek,el]{M.~Bertaina,}
  \author[lf]{P.F.~Bertone,} 
  \author[cb]{S.~Blin-Bondil,}
  \author[lh]{J.~Bogdan,}
  \author[ld]{J.~Burton,}
  \author[ea,eb]{F.~Cafagna,}
  \author[ec,ed]{R.~Caruso,}
  \author[ei,ej,fg]{M.~Casolino,}
  \author[ba]{K.~\v{C}ern\'{y},}
  \author[lf]{M.J.~Christl,}
  \author[ef,eg]{R.~Colalillo,}
  \author[la]{H.J.~Crawford,}
  \author[cb]{A.~Creusot,}
  \author[lm]{A.~Cummings,}
  \author[lc]{J.~Desiato,}
  \author[lb]{R.~Diesing,}
  \author[ef,eg]{A.~Di~Nola,}
  \author[fg]{T.~Ebisuzaki,}
  \author[lb]{J.~Eser,}
  \author[eo]{F.~Fenu,}
  \author[ek,el]{S.~Ferrarese,}
  \author[lb]{G.~Filippatos,}
  \author[lc]{W.W.~Finch,}
  \author[ef,eg]{F.~Flaminio,}
  \author[lb]{S.~Flannery,}
  \author[eq]{C.~Fornaro,}
  \author[lb]{N.~Friedlander,}
  \author[lc]{D.~Fuehne,}
  \author[ja]{C.~Fuglesang,}
  \author[lh]{S.~Gadamsetty,}
  \author[li]{D.~Garg,}
  \author[lh]{E.~Gazda,}
  \author[ek,el]{A.~Golzio,}
  \author[ef,eg]{F.~Guarino,}
  \author[ca]{C.~Gu\'epin,}
  \author[lc]{T.~Heibges,}
  \author[la]{E.G.~Judd,}
  \author[li]{L.~Kupari,}
  \author[ib]{P.A.~Klimov,}
  \author[lj]{J.F.~Krizmanic,}
  \author[lc]{V.~Kungel,}
  \author[ld]{E.~Kuznetsov,}
  \author[ek]{M.~Mignone,}
  \author[ek,el]{M.~Manfrin,}
  \author[ha]{W.~Marsza{\l},}
  \author[lg]{J.N.~Matthews,}
  \author[lb]{K.~Mehling,}
  \author[ef,eg]{M.~Mese,}
  \author[lb]{S.S.~Meyer,}
  \author[ek,el]{H.~Miyamoto,}
  \author[ib]{A.S.~Murashov,}
  \author[li]{J.~M.~Nachtman,}
  \author[lb]{A.V.~Olinto,}
  \author[li]{Y.~Onel,}
  \author[ef]{G.~Osteria,}
  \author[lh]{A.N.~Otte,}
  \author[ef,eg]{B.~Panico,}
  \author[cb,cc]{E.~Parizot,}
  \author[le]{T.~Paul,}
  \author[ba]{M.~Pech,}
  \author[ef] {F.~Perfetto}
  \author[hb]{L.W.~Piotrowski,}
  \author[ei,ej]{Z.~Plebaniak,}
  \author[li]{J.~Posligua,}
  \author[lh]{M.~Potts,}
  \author[cb]{G.~Pr\'ev\^ot,}
  \author[hd]{M.~Przybylak,}
  \author[ld]{P.~Reardon,}
  \author[li]{M.H.~Reno,}
  \author[ee]{M.~Ricci,}
  \author[lh]{O.F.~Romero~Matamala,}
  \author[lc]{F.~Sarazin,}
  \author[bb]{P.~Schov\'{a}nek,}
  \author[ef,eg]{V.~Scotti,}
  \author[ha]{K.~Shinozaki,}
  \author[le]{J.F.~Soriano,}
  \author[lh]{S.~Stephanoff,}
  \author[lc]{P.~Sternberg,}
  \author[lb]{B.K.~Stillwell,}
  \author[hc]{J.~Szabelski,}
  \author[fg]{Y.~Takizawa,}
  \author[cb, ib]{D.~Trofimov,}
  \author[ja]{F.~Unel,}
  \author[eo]{V.~Vagelli,}
  \author[ef,eg]{L.~Valore,}
  \author[lj]{T.~M.~Venters,}
  \author[ld]{J.~Watts~Jr.,}
  \author[lc]{L.~Wiencke,}
  \author[lc]{H.~Wistrand,}
  \author[lf]{R.~Young}

\affiliation[ba]{Joint Laboratory of Optics, Faculty of Science, Palack\'{y} University, Olomouc, Czech Republic}
\affiliation[bb]{Institute of Physics of the Czech Academy of Sciences, Prague, Czech Republic}
\affiliation[ca]{Laboratoire Univers et Particules de Montpellier (LUPM), Universit\'e de Montpellier, CNRS/IN2P3, F-34095 Montpellier, France}
\affiliation[cb]{Universit\'e de Paris, CNRS, AstroParticule et Cosmologie, F-75013 Paris, France}
\affiliation[cc]{Institut Universitaire de France (IUF), France}
 \affiliation[ea]{Istituto Nazionale di Fisica Nucleare - Sezione di Bari, Italy}
 \affiliation[eb]{Universit\`a degli Studi di Bari Aldo Moro, Italy}
 \affiliation[ec]{Dipartimento di Fisica e Astronomia "Ettore Majorana", Universit\`a di Catania, Italy}
 \affiliation[ed]{Istituto Nazionale di Fisica Nucleare - Sezione di Catania, Italy}
 \affiliation[ee]{Istituto Nazionale di Fisica Nucleare - Laboratori Nazionali di Frascati, Italy}
 \affiliation[ef]{Istituto Nazionale di Fisica Nucleare - Sezione di Napoli, Italy}
 \affiliation[eg]{Universit\`a di Napoli Federico II - Dipartimento di Fisica "Ettore Pancini", Italy}
 \affiliation[eh]{INAF - Istituto di Astrofisica Spaziale e Fisica Cosmica di Palermo, Italy}
 \affiliation[ei]{Istituto Nazionale di Fisica Nucleare - Sezione di Roma Tor Vergata, Italy}
 \affiliation[ej]{Universit\`a di Roma Tor Vergata - Dipartimento di Fisica, Roma, Italy}
 \affiliation[ek]{Istituto Nazionale di Fisica Nucleare - Sezione di Torino, Italy}
 \affiliation[el]{Dipartimento di Fisica, Universit\`a di Torino, Italy}
 \affiliation[eo]{ASI (Agenzia Spaziale Italiana)}
 \affiliation[eq]{International Telematic University UNINETTUNO, Rome}
\affiliation[fg]{RIKEN, Wako, Japan}
\affiliation[ha]{National Centre for Nuclear Research, Warsaw, Poland}
\affiliation[hb]{Faculty of Physics, University of Warsaw, Poland}
\affiliation[hc]{Stefan Batory Academy of Applied Sciences, Skierniewice, Poland}
\affiliation[hd]{University of \L{}\'{o}d\'{z}, Poland}
\affiliation[ib]{Skobeltsyn Institute of Nuclear Physics, Lomonosov Moscow State University, Russia}
\affiliation[ja]{KTH Royal Institute of Technology, Stockholm, Sweden}
\affiliation[la]{Space Science Laboratory, University of California, Berkeley, CA, USA}
\affiliation[lb]{University of Chicago, IL, USA}
\affiliation[lc]{Colorado School of Mines, Golden, CO, USA}
\affiliation[ld]{University of Alabama in Huntsville, Huntsville, AL, USA}
\affiliation[le]{Lehman College, City University of New York (CUNY), NY, USA}
\affiliation[lf]{NASA Marshall Space Flight Center, Huntsville, AL, USA}
\affiliation[lg]{University of Utah, Salt Lake City, UT, USA}
\affiliation[lh]{Georgia Institute of Technology, Atlanta, GA, USA}
\affiliation[li]{University of Iowa, Iowa City, IA, USA}
\affiliation[lj]{NASA Goddard Space Flight Center, Greenbelt, MD, USA}
\affiliation[lm]{Pennsylvania State University, PA, USA}

\abstract{ 
The Extreme Universe Space Observatory on a Super Pressure Balloon 2 (EUSO-SPB2) is a pathfinder mission toward a space-based observatory such as the Probe of Extreme Multi-Messenger Astrophysics (POEMMA). The aim of POEMMA is the observation of Ultra High Energy COsmic Rays (UHECRs) in order
to elucidate their nature and origins and to discover $\gtrsim$ 20 PeV very high energy neutrinos that originate from transient and steady astrophysical sources.  EUSO-SPB2 was launched from W\=anaka New Zealand on May 13th, 2023 as a NASA Balloon Program Office test flight. The mission goals included making the 
first near-space altitude observations of the fluorescence emission from UHECR-induced
extensive air showers (EASs) and making the first direct Cherenkov light emission from PeV cosmic rays traversing Earth's atmosphere.  In addition, a Target of Opportunity program
was developed for selecting and scheduling observations of potential neutrino sources as they passed just below the Earth's limb. 
Although a leaky balloon forced termination over the Pacific Ocean after 37 hours, data was collected to demonstrate the successful commissioning and operation of the instruments. This paper includes a description of the payload and the key instruments, pre-flight instrument characterizations in the lab and in the desert, flight operations and examples of the data collected. The flight was too short to catch a UHECR event via fluorescence, however about 10 candidate EAS events from cosmic rays were recorded via Cherenkov light.  
}


\arxivnumber{xxxx.yyyyy} 

\begin{document}
\maketitle
\flushbottom

\input{1_Mission-Science-Goals}
\input{2_Payload-Overview}

\input{3_Gondola-System}
\input{4_Opto_Mech}

\input{5_FT}

\clearpage
\input{6_CT}
\input{7_UCIRC}
\input{9_Flight}

\input{99_Conclusions}

\newpage
\acknowledgments
This work was partially supported by Basic Science Interdisciplinary Research Projects of RIKEN and JSPS KAKENHI Grant (22340063, 23340081, and 24244042), by the ASI-INAF agreement n.2017-14-H.O, by the Italian Ministry of Foreign Affairs and International Cooperation, by the Italian Space Agency through the ASI INFN agreements Mini-EUSO n.2016-1-U.0, n. 2017-8-H.0, OBP (n. 2020-26-Hh.0) and n. 2021-8-HH.0. by NASA award 11-APRA-21730058, 16-APROBES16-0023, 17-APRA17-0066, NNX17AJ82G, NNX13AH54G, 80NSSC18K0246, 80NSSC18K0473, 80NSSC19K0626, 80NSSC18K0464, 80NSSC22k0426 and 80NSSC19K0627in the USA, by the French space agency CNES, by the Deutsches Zentrum f\"ur Luft- und Raumfahrt, the Helmholtz Alliance for Astroparticle Physics funded by the Initiative and Networking Fund of the Helmholtz Association (Germany), by Slovak Academy of Sciences MVTS JEM–EUSO, by National Science Centre in Poland grants 2017/27/B/ST9/02162 and 2020/37/B/ST9/01821, by Deutsche Forschungsgemeinschaft (DFG, German Research Foundation) under Germany Excellence Strategy -EXC-2094-390783311, by Mexican funding agencies PAPIIT-UNAM, CONACyT and the Mexican Space Agency (AEM), as well as VEGA grant agency project 2/0132/17, and by State Space Corporation ROSCOSMOS and the Interdisciplinary Scientific and Educational School of Moscow University ``Fundamental and Applied Space Research".

\bibliographystyle{JHEP}
\bibliography{References.bib}

\end{document}

%% file: 1_Mission-Science-Goals.tex
\section{Mission Science Goals}\label{sec:Mission-Science-Goals}

Ultra High Energy Cosmic Rays (UHECRs) are the most energetic subatomic particles ever observed, with energies reaching $10^{20}{\rm~eV}$ (100 EeV) and above. Although the existence of UHECRs has been established for decades by ground-based observatories, their
sources and underlying acceleration mechanisms remain uncertain~\cite{Kotera:2011cp, AlvesBatista:2019tlv}. Measurements from the Pierre Auger Observatory~\cite{PierreAuger:2017pzq} favor an extragalactic origin for UHECR.

A multimessenger window to the most extreme energetic environments in the universe may be provided by the combination of observations of UHECRs, 
astrophysical PeV neutrinos, gamma rays, and gravitational waves. The emergence of multi-messenger astronomy
is intriguing. For example, in 2017 the IceCube Neutrino Observatory provided an alert concerning a 270 TeV neutrino to the astrophysics community, and observations of gamma rays by Fermi-LAT and MAGIC corresponded to the IceCube event direction with a chance probability of $3 \sigma$, indicating probable multi-messenger observation of a flaring blazar~\cite{IceCube:2018dnn}. More recently, IceCube has reported a neutrino excess associated with the nearby active galaxy NGC 1068 with a $4.2\sigma$ significance~\cite{IceCube:2022der}.

Space-based observation of UHECRs was proposed in the 1980s~\cite{Benson:1981xe} as a way to achieve uniform coverage of the entire celestial sphere and to image a large volume of observable atmosphere in which UHECRs initiate particle cascades. More recently, several pathfinder missions designed to establish the technologies and techniques for space-based observations have been
performed or are in preparation as part of the Joint Exploratory Missions for an Extreme Universe Observatory (JEM-EUSO)~\cite{JEM-EUSO}. Taken together, these complementary missions form stepping stones to a large
space-based observatory, such as the Probe of Extreme Energy Multi-Messenger Astrophysics (POEMMA)~\cite{POEMMA:2020ykm}.
In this article, we describe the Extreme Universe Space Observatory on a Super Pressure Balloon 2 (EUSO-SPB2) mission, a balloon-borne instrument designed to explore the
scientific and technical aspects of POEMMA. 

The principal scientific objective of POEMMA is the elucidation of the origins and nature of UHECRs~\cite{Anchordoqui:2019omw}
and to discover $E \gtrsim 20~\rm{PeV}$ very high energy neutrinos that originate from transient and steady astrophysical sources~\cite{transients}. The POEMMA mission design envisions a pair of identical satellites that fly in formation to make
coincident measurements of light emission from Extensive Air Showers (EASs). Each satellite comprises a Schmidt 
telescope and a hybrid focal surface, most of which is instrumented with Multi-Anode Photo Multiplier Tubes
(MAPMTs) with the remainder employing Silicon Photomultipliers (SiPMs). 

For UHECR-induced EASs, the MAPMTs measure fluorescence light emitted by atmospheric nitrogen
excited by the EASs with 1$\mu s$ time resolution. Stereo observation by two telescopes
allows for superior geometrical reconstruction, leading to good resolution on energy and the
mass-sensitive parameter $X_{\rm max}$, the depth at which the shower
reaches its maximum size in terms of particle number. Following a stereo requirement that the angle between the two planes defined by the detectors and the shower axis be at least $5^{\circ}$ in this study, the expected energy resolution at $5 \times 10^{19}{\rm~eV}$ is 20\%, the $X_{\rm max}$ resolution is 30 $\rm{g}/\rm{cm}^2$ and the angular resolution is $1.5^{\circ}$. Together, these measurements allow for composition-enhanced anisotropy measurements over the full celestial sphere.

Upward-traveling EASs can be produced by PeV $\nu_\tau$ interactions in the Earth's limb leading to the emergence of a $\tau$ that decays in the atmosphere. Such showers can be observed by the brief flash
of forward-beamed Cherenkov light produced by the EAS~\cite{PhysRevD.103.043017}. Both satellites will be able to detect this
signal by exploiting the 20 ns time resolution of the SiPMs, which is sufficient to disentangle background light~\cite{PhysRevD.100.063010,PhysRevD.102.123013}. The satellites can be rotated away from baseline nadir pointing to point the center of the 
SiPM Field of View (FoV) in the directions of potential astrophysical transient sources behind the Earth's limb. Observations of such sources are referred to as Target-of-Opportunity (ToO) Observations~\cite{Reno:2021xos, transients}.

EUSO-SPB2 was designed to benchmark the detection techniques proposed for POEMMA by flying two
Schmidt telescopes. They were flown on a Super Pressure Balloon (SPB) that floated at 33 km~\cite{Cummings:2023srw} on an Ultra-Long Duration Balloon (ULDB) as a mission of opportunity launched from NASA's mid-latitude W\=anaka, New Zealand facility. The EUSO-SPB2 mission builds on the 2017 EUSO-SPB1 mission \cite{ABDELLAOUI2024102891} launched from W\=anaka and the 2014 EUSO-Balloon mission \cite{Abdellaoui_2018,Adams:2022oko} launched from Timmins Canada.  

\begin{figure}[!h]
\centering
\includegraphics[width=1.\textwidth]{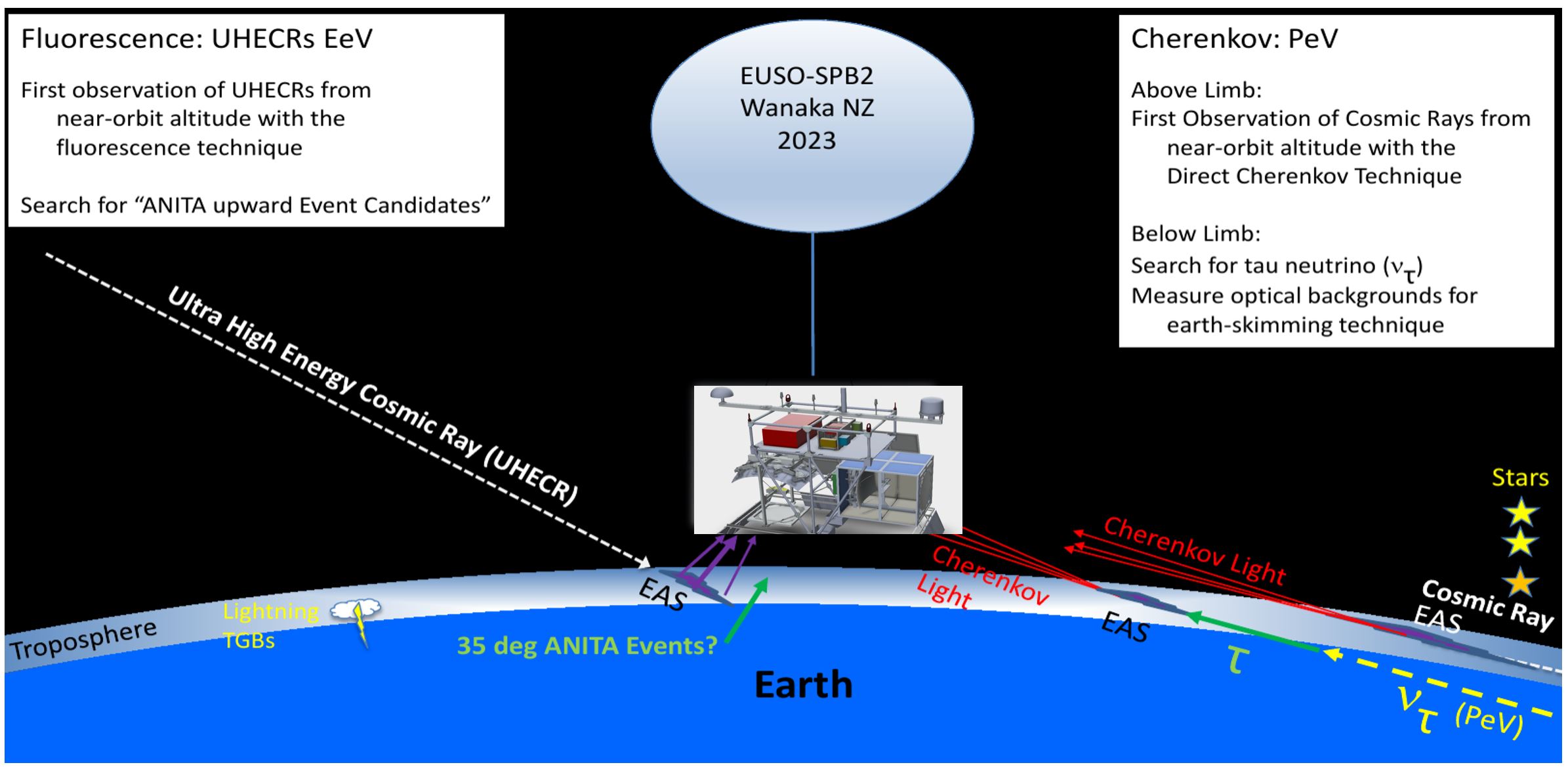}
\caption{Illustration of the EUSO-SPB2 mission goals.}
\label{fig:science_diagram}
\end{figure}

For EUSO-SPB2, one telescope was equipped with MAPMTs with fixed nadir pointing for detecting EAS fluorescence signals. The second telescope was equipped
with SiPMs for observation of direct Cherenkov light and was capable of pointing both above and below
Earth's limb. The mission goals included making the
first near space altitude observations of fluorescence emission from UHECRs and direct Cherenkov emission from PeV cosmic rays.~\cite{ICRC2021_EUSO-SPB2Overview, Wiencke:2019vke, Filippatos_2021}. In addition, a ToO program
was developed~\cite{Heibges:2023yhn,  Posligua:2023cdm} to observe targets below the limb. The configuration and mission goals are shown in Figure~\ref{fig:science_diagram}.

%% file: 2_Payload-Overview.tex
\section{Payload Overview}\label{sec:Payload-Overview}

\begin{figure}[!ht]
\centering
\includegraphics[width=1.\textwidth]{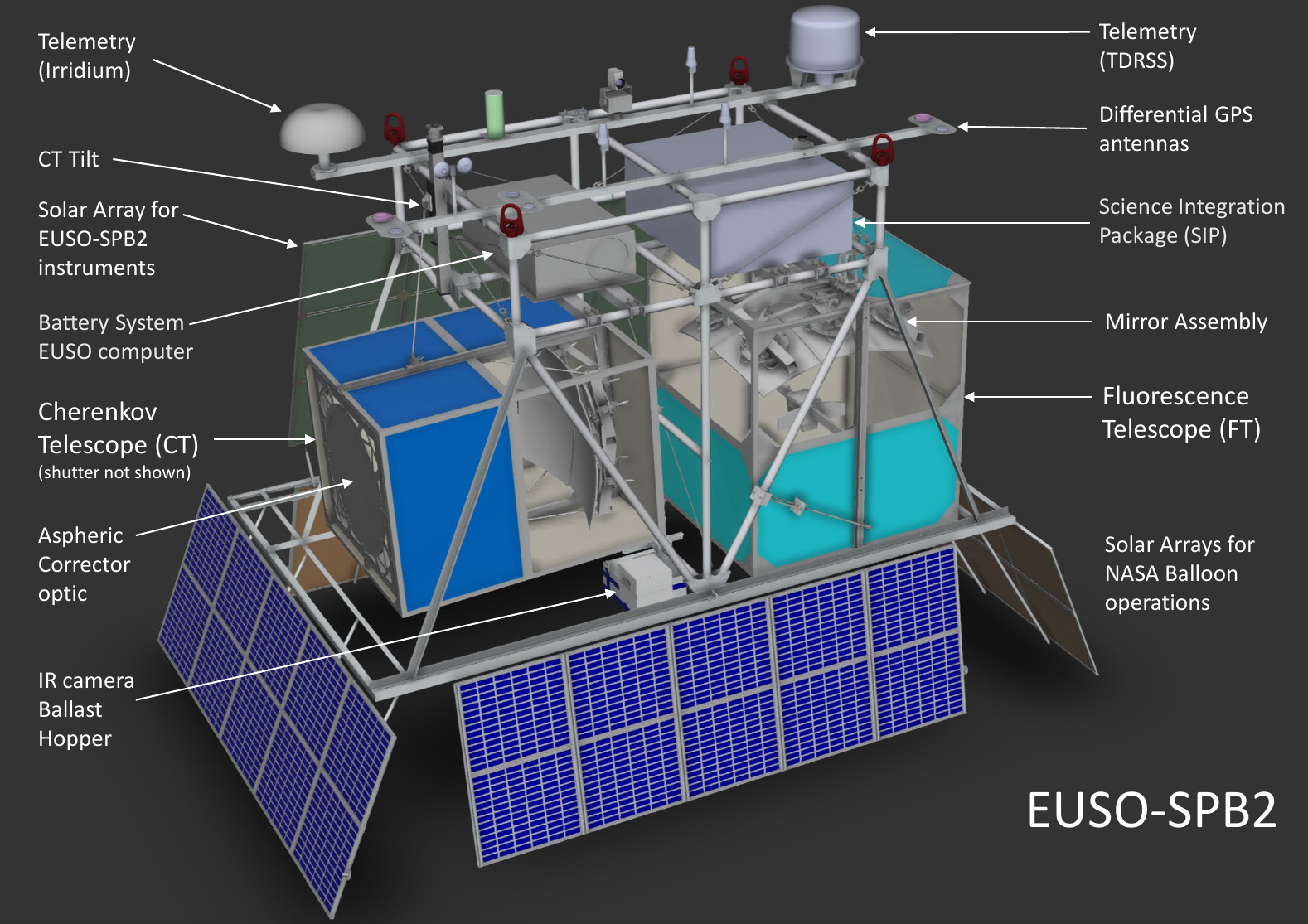}
\caption{ The payload arrangement shows the two telescopes (FT and CT) and the IR cloud camera, as well as the long-duration balloon flight gear, solar panels and telemetry antennas. The Starlink antenna is not shown. Several telescope enclosure panels were removed in this rendering.}
\label{fig:payload}
\end{figure}

The EUSO-SPB2 payload, illustrated in Figure \ref{fig:payload}, featured an optical Cherenkov Telescope (CT) and a Fluorescence Telescope (FT), their associated control and readout systems, the second generation of the University of Chicago Infrared Camera (UCIRC2) to characterize clouds below the FT. The CT included aperture shutters and a tilting system. The mission support equipment flown onboard included two solar power systems designed to support a long duration flight, NASA telemetry systems, a control computer for routing commands from the ground to the CT, FT and IR Camera and for storing and routing data to the ground. The gondola also carried NASA equipment for operating and monitoring components of the balloon system, the NASA Science Interface Package (SIP), an independent Starlink\texttrademark~telemetry system, and 600 lbs of releasable ballast. The parameters of the instruments, the solar power system, the balloon and the flight are summarized in Table~\ref{tab:spec}.

The payload was required to meet two weight limit requirements. There was a 3000 lb limit for the ''science" components on the payload that included the gondola structure. There was an overall limit of 5500 lbs for the total weight suspended from the balloon. The EUSO-SPB2 payload was also required to meet various safety, structural and logistic requirements. Many of these are specified in \cite{CSBF:2019aa}.

\begin{table}[htp]
\centering
\begin{tabular}{| l l l l|}
\hline
Item & Specification & Notes&\\
\hline
Fluorescence Telescope & & &\\
\hline
Energy Max Sensitivity & ${\approx}$3~EeV & & \\
Trigger aperture & ${\approx}$50~km$^{2}$sr (5~EeV) &At 33 km altitude &\\
 &${\approx}$500~km$^{2}$sr (10~EeV) &  & \\
Pointing Direction & Nadir & 6 segment spherical mirror &\\
Entrance pupil & 1 m diameter & PMMA Corrector &\\
Field of view & 3${\times}$12\textdegree${\times}$12\textdegree & &\\
Pixel field of view & 0.2\textdegree${\times}$0.2\textdegree & For active area&\\
Pixel ground footprint & 120~m${\times}$120~m & As projected from 33 km&\\
Number of pixels & 6912 (3${\times}$48${\times}$48)& 108~MAPMTs${\times}$64 pixels each&\\
MAPMT & R11265-203-M64~& Hamamatsu &\\
UV transmitting filter & BG-3,~2~mm thick& 1 per PDM (36 MAPMTs) &\\
Readout & DC coupled & 6~ns double-pulse resolution &\\
Time-bin duration & 1.0~${\mu}$s integration & Event packet 128~bins (128~${\mu}$s)&\\
\hline
Cherenkov Telescope & & &\\
\hline
Energy threshold & ${\approx}$1~PeV & & \\
Tilting of optical axis &+0.2\textdegree ~to -10.1\textdegree  & Relative to horizontal&\\
Telescope optics & Modified Schmidt& 4 segment spherical mirror &\\
Entrance pupil & 1 m diameter & PMMA Corrector &\\
Field of view & 6.4\textdegree(V)${\times}$12.8\textdegree(H) & &\\
Pixel field of view & 0.4\textdegree${\times}$0.4\textdegree & Pixel Size: 6~mm${\times}$6~mm &\\
Number of pixels & 512 (16V${\times}$32H)& 32~SiPMs of 4${\times}$4 pixels each&\\
SiPM & S14521-6050AN-04~& Hamamatsu &\\
Time-bin duration & 10~ns integration & Event packet 512~bins (5.1~${\mu}$s)&\\
\hline
Balloon & 0.5${\times}10^{6}$ m$^{3}$~(18${\times}10^{6}$ ft$^{3}$) &Helium & \\
Nominal float height & 33.5 km (110000~ft) & &\\
Telemetry (data) & ${\approx}$200 ~Mbits~s$^{-1}$& 1~Starlink (maritime unit)& \\
                 & ${\approx}$75~kbits~s$^{-1}$& 1~TDRSS& \\
                 & ${\approx}$75~kbits~s$^{-1}$& 1~Iridium OpenPort& \\
Telemetry (comms) & ${\approx}$1.2~kbits~s$^{-1}$ &2~Iridium Pilots& \\
Power consumption & 200/420~W (day/night)& w/ battery heater at night&\\
Batteries &6${\times}$24 A${\cdot}$h Lithium-Ion &Valence U27-24XP&\\
Solar panels & 15${\times}$100~W  & SunCat Solar &\\
Detector weight & 1223~kg (2250~lb) &w/o SIP, antennas, and ballast & \\
Releasable ballast & 272~kg (600~lb)& 0~kg remaining at termination & \\
Total weight & 2557~kg (5625~lb)&Everything below balloon&\\
\hline
Flight start &2023 May 13 00:02~UT ~& 44.7218\textdegree S 169.2540\textdegree E&\\
Flight end &2023 May 14 12:54~UT ~& 34.0831\textdegree S 151.8768\textdegree W& \\
Flight duration &  36~hr 52~mn &Leaky Balloon& \\

\hline
\end{tabular}
\caption{EUSO-SPB2 payload and mission specifications}
\label{tab:spec}
\end{table}

%% file: 3_Gondola-System.tex
\section{Gondola System}\label{sec:Gondola-System}
The EUSO-SPB2 gondola system included a mechanical assembly to which everything on the payload was attached. It also included the subsystems that supported the operation of the CT, FT, IR camera, telemetry systems, solar power systems, GPS systems, and the gondola control computer system.

\begin{figure}[!ht]
\centering
\includegraphics[width=0.75\textwidth]{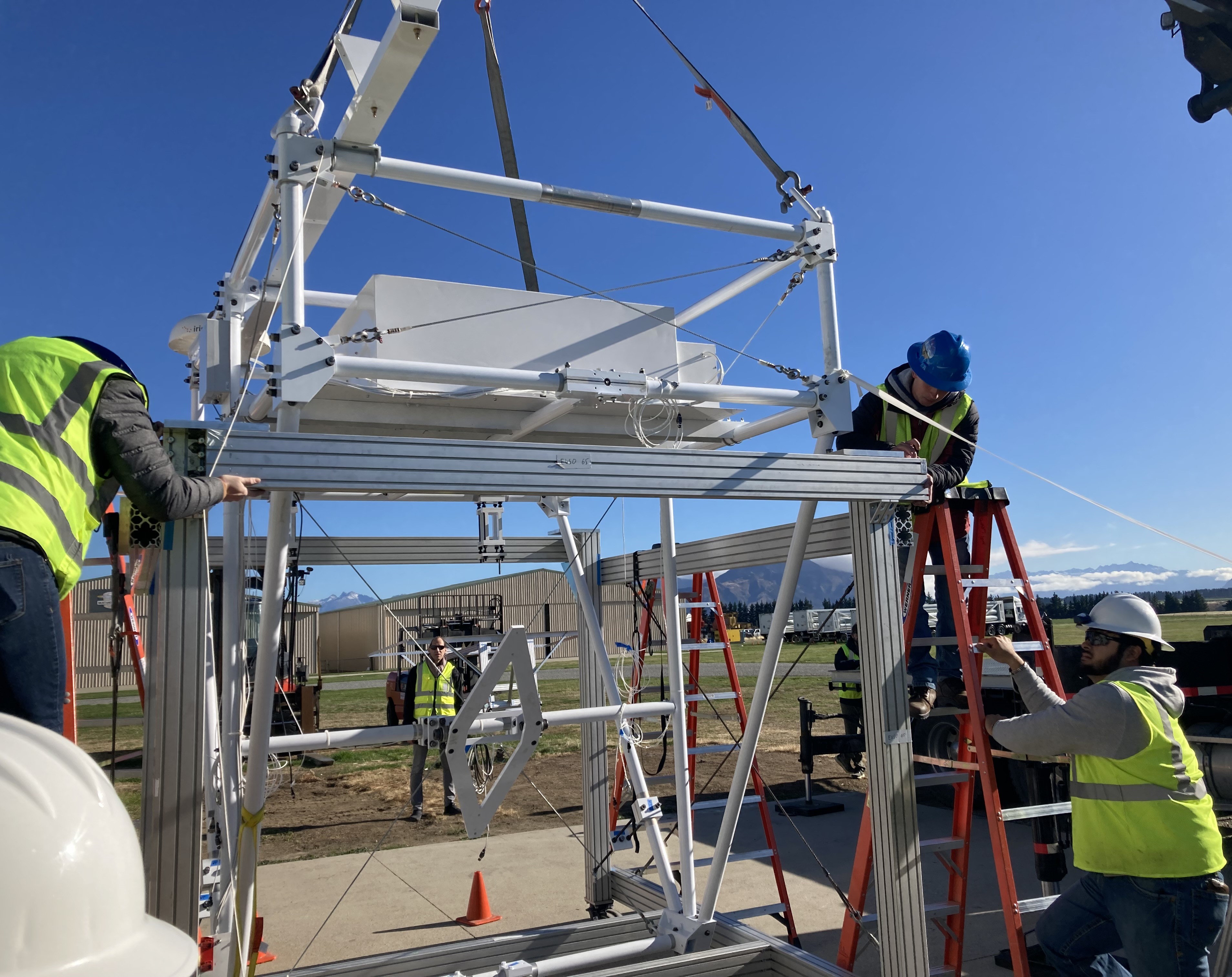}
\caption{ The gondola structure is lowered onto the payload roll-out cart. The box on top contains the batteries and gondola control computer. The back of the Cherenkov telescope bolts to the diamond-shaped mounting bracket that can pivot up and down. The roll-out cart does not fly.}
\label{fig:gondola_structure}
\end{figure}

\subsection{Mechanical Assembly}
To accommodate the required viewing directions of the two telescopes and to save weight, a wedge-shaped gondola structure (Figure \ref{fig:gondola_structure}) was constructed. In contrast to the high-g vibrational forces during a rocket launch, a balloon launch is relatively benign. The major mechanical shocks that drive the structural requirements of the gondola occur during the normal flight termination sequence. The balloon is deflated quickly. The abrupt loss of lift suddenly releases tension in the flight train and this generates a shock on the gondola. Then the parachute opens, which exerts a second set of bounces. The distribution of these dynamic forces between 4 attachment points at the top of the gondola is unpredictable. The essence of the applicable safety requirements is that the gondola remains intact and parts do not fall off the payload during this sequence. The gondola must handle the worst-case scenario in which the maximum dynamic load is held by any one, and exactly one, of the attachment points. For the bolted joints used in our design, the gondola must be able to survive an 8-g vertical~$\pm~45\degree$ load and a 4-g transverse load exerted at any of the 4 attachment points. This corresponded to 197,000 N (44,000 lbf). To achieve this, the mechanical design of the gondola was modeled through extensive finite element analysis that propagated this worst-case loading through the structure. High-strength 7075 aluminum alloy was used for all machined joint blocks and all structural tubing to meet the strength requirement and to remain below the payload mass limit.

\subsection{Telemetry} Four telemetry links mounted on the gondola handled commanding and data downloads. The low-rate iridium pilot links carried commands. Commands could take several minutes and were limited to 256 bytes. Data handling was supported by an Iridium OpenPort link and by a Tracking and Data Relay Satellite System (TDRSS) link. The effective rate for these systems is about 75 kbs. 

Approximately 5 weeks before launch, NASA/CSBF delivered two maritime Starlink units to W\=anaka for the two flights of the 2023 campaign. Starlink is a low-Earth orbit satellite constellation that delivers high-speed, low-latency internet. 
This campaign was the first to fly Starlink units on NASA Ultra Long Duration Balloon (ULDB) flights. In EUSO-SPB2, the unit was isolated from NASA equipment critical to balloon operations by connecting it directly to the FT subnet. A key element of the implementation was the use of reverse SSH tunneling to accommodate unpredictable changes to the Starlink unit's IP address. This Starlink system turned out to be critical for the mission. 

\subsection{Solar Power System} Two independent solar power systems were flown with their own solar panels, battery banks, and charge controller systems to support a mid-latitude ULDB flight.

The balloon power system featured the 4-sided array of panels below the gondola that powered the NASA/CSBF telemetry, rotator, and instrumentation to monitor and operate the balloon. During daylight, at least one side would always point at the Sun for charging even if the NASA rotator failed. This helped ensure that sufficient battery power would be available for critical balloon operations such as GPS tracking, ballast drops, and termination. 

The science solar power system was designed and built by the EUSO-SPB2 collaboration to power the CT, FT, UCIRC2, Gondola Control Computer (GCC) and related systems. The estimated power requirements and the remaining space on the gondola precluded the use of a second 4-sided array. Instead, a single large array of 15 100-watt panels was mounted on one side of the gondola with solar tracking provided by the NASA rotator system during the day. The specifications of the system are listed in Table~\ref{tab:spec}. At first glance, the capacity of the solar panels and the ${24~V}$ battery bank may appear oversized relative to the day and night power draw of the instruments connected to it. However, the power system was designed to support an ULDB flight that spanned the winter solstice (June 21, 2023), together with the possibility that the balloon could drift to higher southern latitudes, where the daytime charging periods would shorten and night observing times would lengthen, for example, yielding an opportunity to collect more science data. A publication describing this system in detail is in preparation~\cite{EUSO:SolarPower:2024a}.

\subsection{Global Positioning System} Multiple GPS systems were flown. Those used to support the EUSO-SPB2 science instruments are noted here. The FT electronics included two differential GPS systems to track the rotational angle of the payload. NASA/CSBF also flew a differential GPS system to provide feedback to the NASA rotator at night to support CT ToO operations. The Gondola Control Computer (GCC) also had its own GPS system. 

\subsection{Gondola Control Computer}
The GCC performed several critical functions. It logged and relayed commands to the CT, FT, Shutter/Tilt, Solar Power, and IR Camera systems that arrived from the ground through the on-board NASA Support Instrumentation Package (SIP). The GCC also logged and relayed the status bytes returned from the EUSO-SPB2 systems. The GCC also relayed data from these systems to the ground using an automated priority scheme following a bandwidth allocation between the different subsystems. As part of the GCC development, a GCC virtual machine (GCCVM) was created. Subsystem experts installed the GCCVM on laptops at their home institutions to develop and test command sequences on their subsystems before delivering them for telescope and payload integration.

The GCC was powered through a 8-channel Power Distribution Unit (PDU) that routed DC power from the battery bank to the various EUSO-SPB2 subsystems. Since the GCC also controlled the PDU power switching, the PDU's external reset line was interfaced to an output line on the SIP. In the event the GCC got into an unresponsive state, this line could be toggled remotely to power-cycle the entire EUSO-SPB2 system including the GCC. This complete reset was demonstrated during the payload integration and check-out on site.

The GCC was housed in a sealed cylindrical vessel pressurized at 1 atmosphere. This vessel was located in a box on the top of the gondola that also housed the battery bank for the solar power system.

%% file: 4_Opto_Mech.tex
\section{Telescope Opto-Mechanics} \label{sec:Optics}
Since the optical and mechanical designs of the FT and CT are quite similar, they are described together.

\begin{figure}[!ht]
\centering
\includegraphics[width=0.9\textwidth]{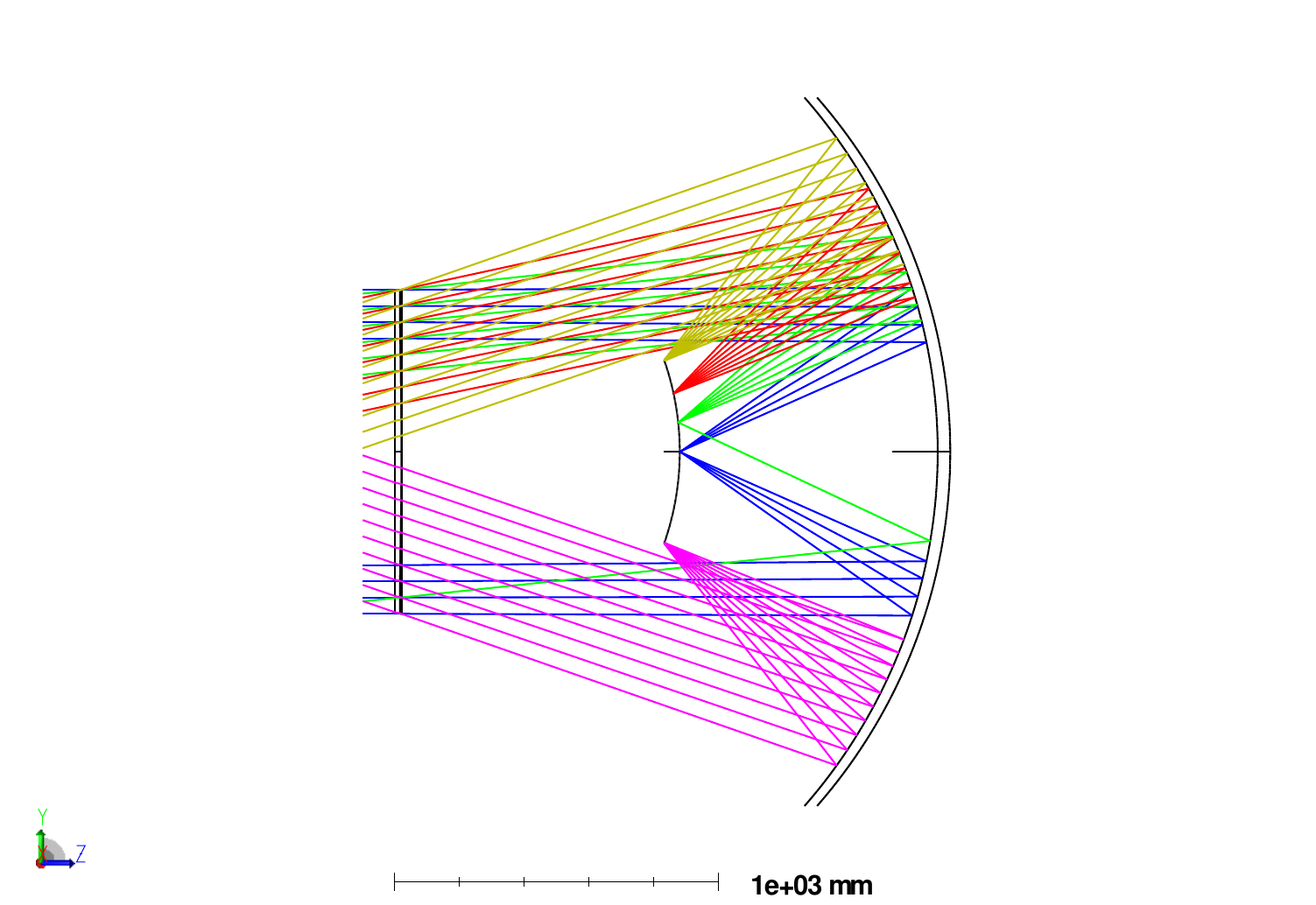}
\caption{ The EUSO-SPB2 optics design is illustrated in this top-down view. From left to right are featured a 1 m diameter entrance pupil occupied by an Aspheric Corrector Plate, a curved image surface, and a spherical primary mirror.}
\label{fig:optics_design}
\end{figure}
\subsection{Design \& Fabrication}
The modified Schmidt optical design of the CT and FT features identical 1 m diameter entrance pupils, spherical segmented primary mirrors with the same 1.6 m radii of curvature and identical curved image surface shapes as illustrated in Figure \ref{fig:optics_design}. To span the larger FoV of the FT, the FT primary mirror was constructed from 6 mirror segments. The smaller FoV of the CT could be captured using a four-segment primary mirror. The 10 mirror segments were identical and interchangeable between telescopes.

The entrance pupils were covered by an aspheric corrector plate (ACP) optic fabricated from UV-transmitting Polymethyl Methacrylate (PMMA). The ACPs were flat on the surface facing away from the telescope and curved on the other side. The fabrication process used a high-precision diamond turning process followed by polishing. 

The borosilicate 3.3 glass mirror segments were vacuum-slumped on a custom mold in an oven followed by polishing and then cut to the required dimensions. The front surface of the mirror segments was coated with a thin film of aluminum using a vacuum evaporation process followed by a protective silicon dioxide coating.  The thickness of the coatings was selected to optimize reflectivity in the UV. The reflectivity as a function of wavelength is presented in Section \ref{FT}. The backs of the mirror segments were roughened to facilitate the process of bonding them to a metal mounting structure.

This modular structure was designed to hold the mirror segments subject to the following conditions:
\begin{itemize}
    \item adapt to the shape of the glass without introducing stress on the glass,
    \item allow for manual adjustments to tilt each segment in two orthogonal direction, translate individual segments back and forth, rotate (clock) each segment,
    \item survive a 12-g load in stress or sheer over a temperature range of +45\degree C to~-50\degree C and
    \item be as light-weight as reasonably achievable.
\end{itemize}
The mirror segment mounting assembly featured three 3-point whiffletree flexure fixtures (Figure \ref{fig:optics_bonding} panels A, B) that provided a total of 9 1" diameter bonding pads that were glued to the back of a mirror segment. The three whiffletree fixtures were attached through a single-point flexure fixture to a circular mounting plate. This geometry of one large triangle holding three smaller triangles in combination with spherical bearings at all 12 junction points provided sufficient toggling to position all 9 bonding pads snugly against the back of a mirror segment. Each of the 12 spherical bearings was supported by three flexures machined into the aluminum. These flexures act as springs to accommodate thermal expansion/contraction effects and also cushion mechanical shocks in stress and sheer. Extensive finite element analysis was part of the design process.

\subsection{Mirror Bonding}
\begin{figure}[!ht]
\centering
\includegraphics[width=1.0\textwidth]{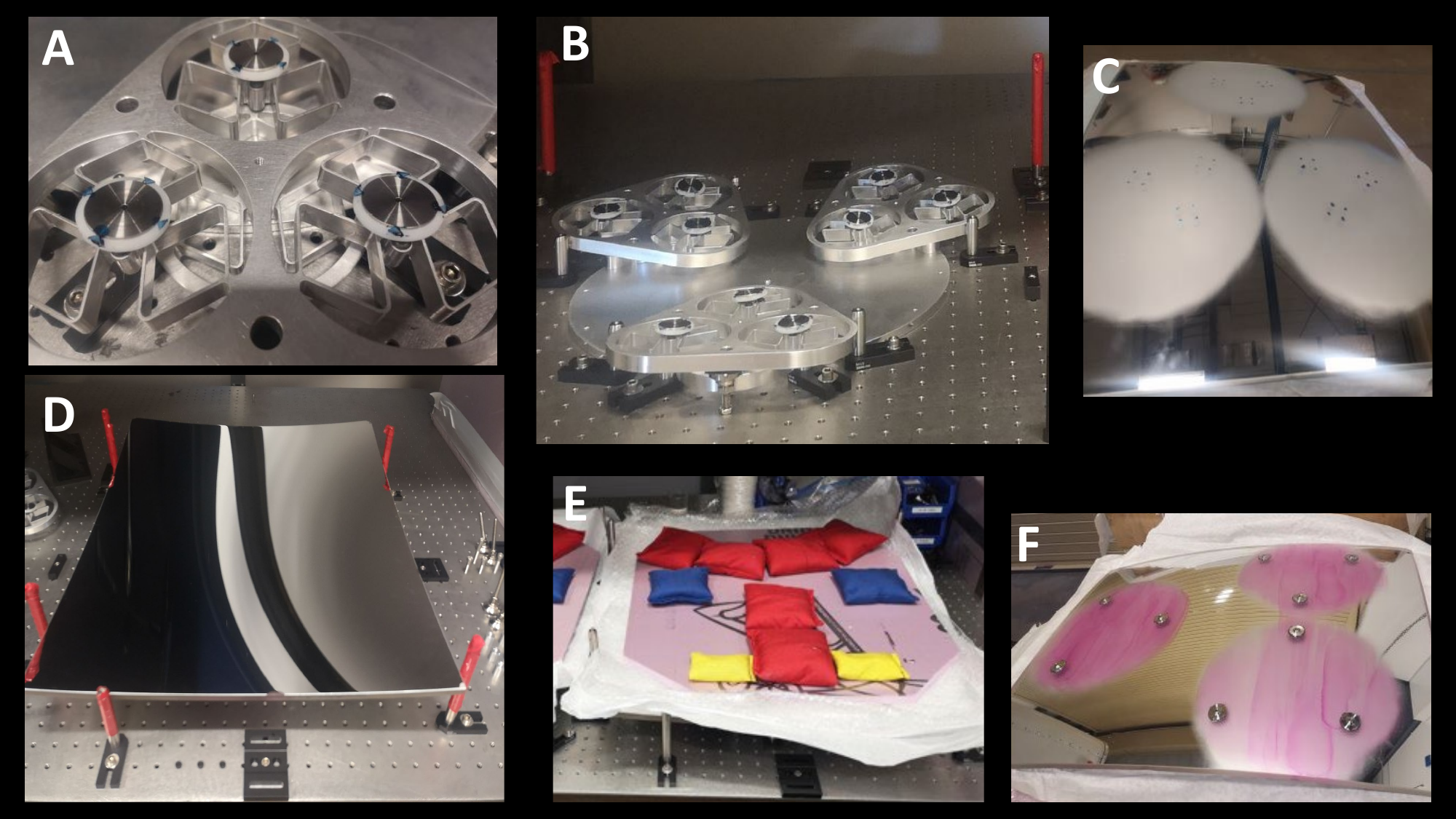}
\caption{Mirror segment mounting and bonding. Panel A: A 3-point whiffletree assembly including 9 diamond-shaped flexures machined into the piece which is shown with three 1" diameter kovar bonding pads attached. The white plastic spacers set the bond-line thickness and are removed for flight. Panel B: Three whiffletree assemblies attached to the circular mounting plate and positioned for bonding. Panel C: The back of one mirror segment showing the roughened regions for bonding. Panel D: A mirror segment is indexed on top of the 9 pads after primer and adhesive have been applied. Panel E: The mirror segment is covered with protective tissue and weighted with bags of lead shot for curing. Panel F: The 9 kovar pads have been bonded to the mirror segment and the mounting assembly has been removed.}
\label{fig:optics_bonding}
\end{figure}

\begin{figure}[!ht]
\centering
\includegraphics[width=0.75\textwidth]{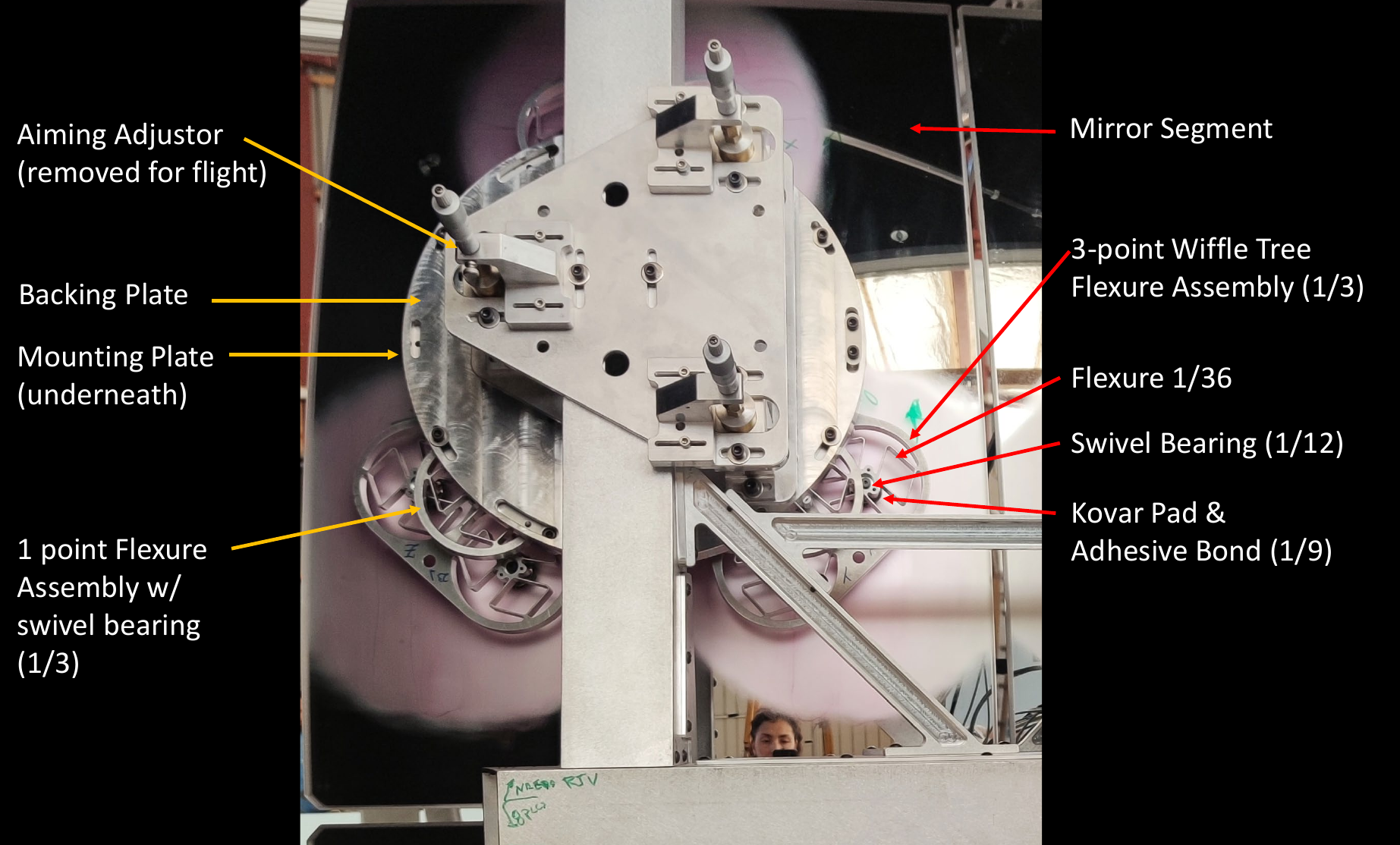}
\caption{A glass mirror segment with mounting hardware is shown attached to the FT. Ten of these assemblies were flown, 6 on the FT and 4 on the CT.}
\label{fig:optics_mirror_mounted}
\end{figure}

Special attention was given to the method used to bond the borosilicate glass mirror segments to their metal mounting structures. Bonding pads of ${\sim2.5~\text{cm}~(1.0")}$ diameter were fabricated from Kovar, a nickel-cobalt alloy engineered to have a coefficient of thermal expansion (CTE) similar to the CTE of borosilicate glass. To maintain a uniform thickness of the bond, the mating surface of each bonding pad was milled to the radius of curvature of the back of the glass. A 2-part epoxy, 3M scotch-weld EA-2216, was selected. After cleaning, the glass surfaces were coated with structural adhesive primer EC-3901. The steps in the process are illustrated in Figure~\ref{fig:optics_bonding}. 
More than 20 sample bonds were prepared and tested over temperature ranges of +45\degree to -50\degree C with a loading equivalent of 12-g. Testing at 12-g was performed in stress (perpendicular) and in sheer (parallel) directions because the FT points down and the CT points sideways. In testing prototype joints with bonding surfaces smaller than the one flown, the warm sheer case was found to be the weakest. Hour-long tests of the bond design used in flight were found to hold 16-g in stress and in sheer at 25\degree C, a temperature well above the warmest temperatures that the mirrors experienced during the 2023 flight. The breaking strength of the bonds above 16-g was not explored. Details are provided in \cite{Kungel:2023Thesis}. A mirror assembly mounted on the FT is illustrated in Figure \ref{fig:optics_mirror_mounted}.
.

\subsection{Laboratory Tests}
\begin{figure}[!ht]
\centering
\includegraphics[width=0.8\textwidth]{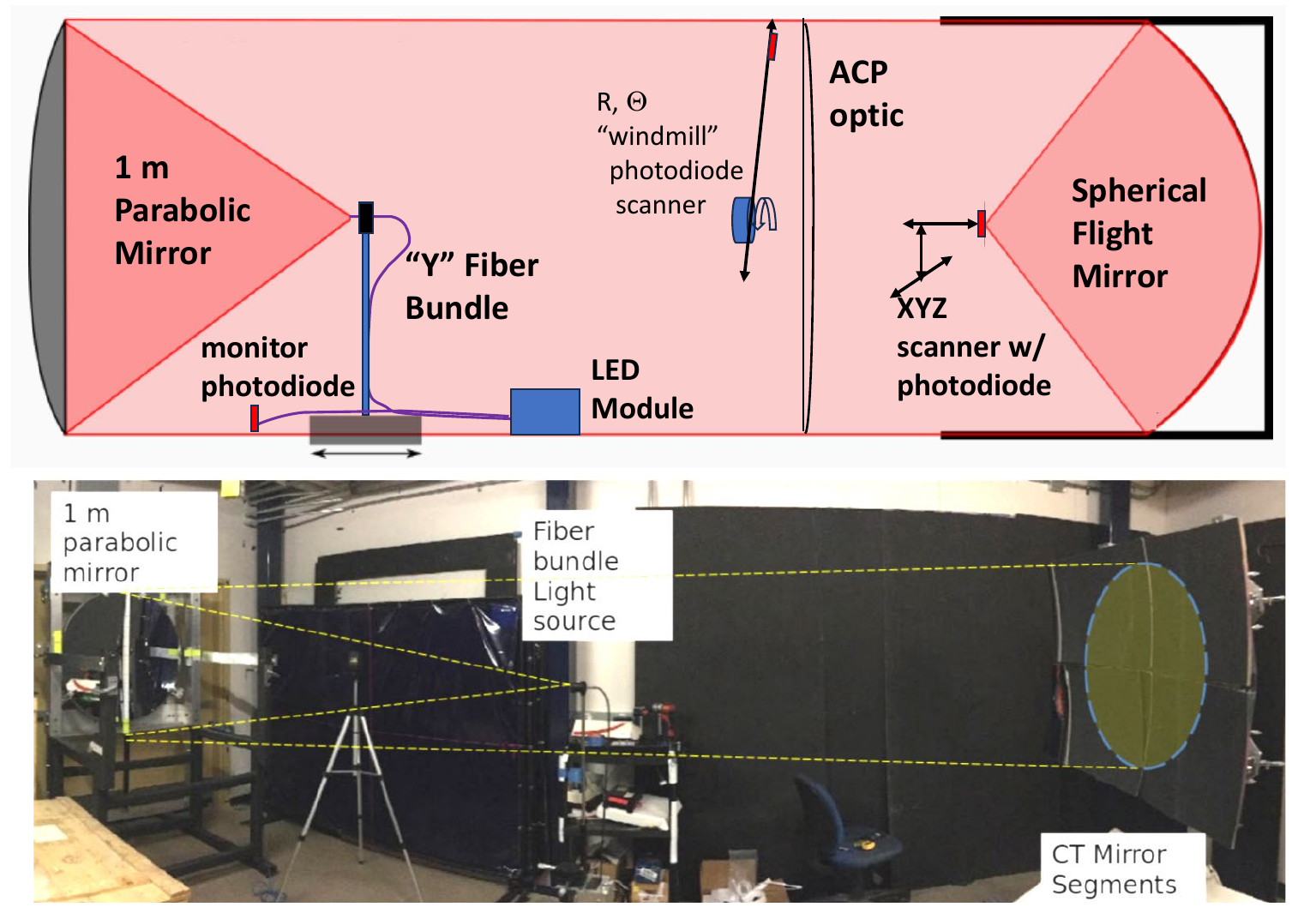}
\caption{ The 1-m parallel test beam setup, (top panel) and in the laboratory (bottom panel). A parabolic mirror delivers a parallel beam to fills the aperture of the telescope. 
plane. The projected beam in the lower picture is not to scale}
\label{fig:optics_1m_beam}
\end{figure}

A 1~m diameter optical test beam system was configured to measure the point spread function (PSF) and the optical efficiency of the telescope optics. Although this beam was much less parallel than starlight, for example, it was sufficiently parallel for this application, conveniently available 24/7, and matched the 1~m diameter of the telescopes' entrance pupils. The arrangement is illustrated in
Figure~\ref{fig:optics_1m_beam}. The beam was produced by positioning one leg of a Y fiber bundle at the focus of an f/3.5 1~m diameter parabolic mirror. The common end of the fiber was driven by an OTS DC LED system. A photodiode was used on the second leg of the Y for monitoring. The parabolic mirror reflects the light toward the flight optics. The test beam can be made parallel at the level of ${\pm 0.01\degree}$. This value is reasonably small compared to the ${0.2\degree \times 0.2\degree}$ pixel FOV of the FT and the ${0.4\degree \times 0.4\degree}$ pixel FOV of the CT. 

A square 1~cm$^{2}$ photodiode with a 0.5~mm square aperture was mounted on an XYZ mechanical scanner that was positioned near the focus of the flight mirror system to be tested. To enable an optical efficiency measurement, a third 1~cm$^{2}$ photodiode was mounted on a ${R/ \theta}$ "windmill" scanner to measure the light incident across the telescope entrance pupil. 
The PSF and optical efficiency measurements for both telescopes are listed in Table~\ref{tab:optics_tests}.

\begin{figure}[!ht]
\centering
\includegraphics[width=0.8\textwidth]{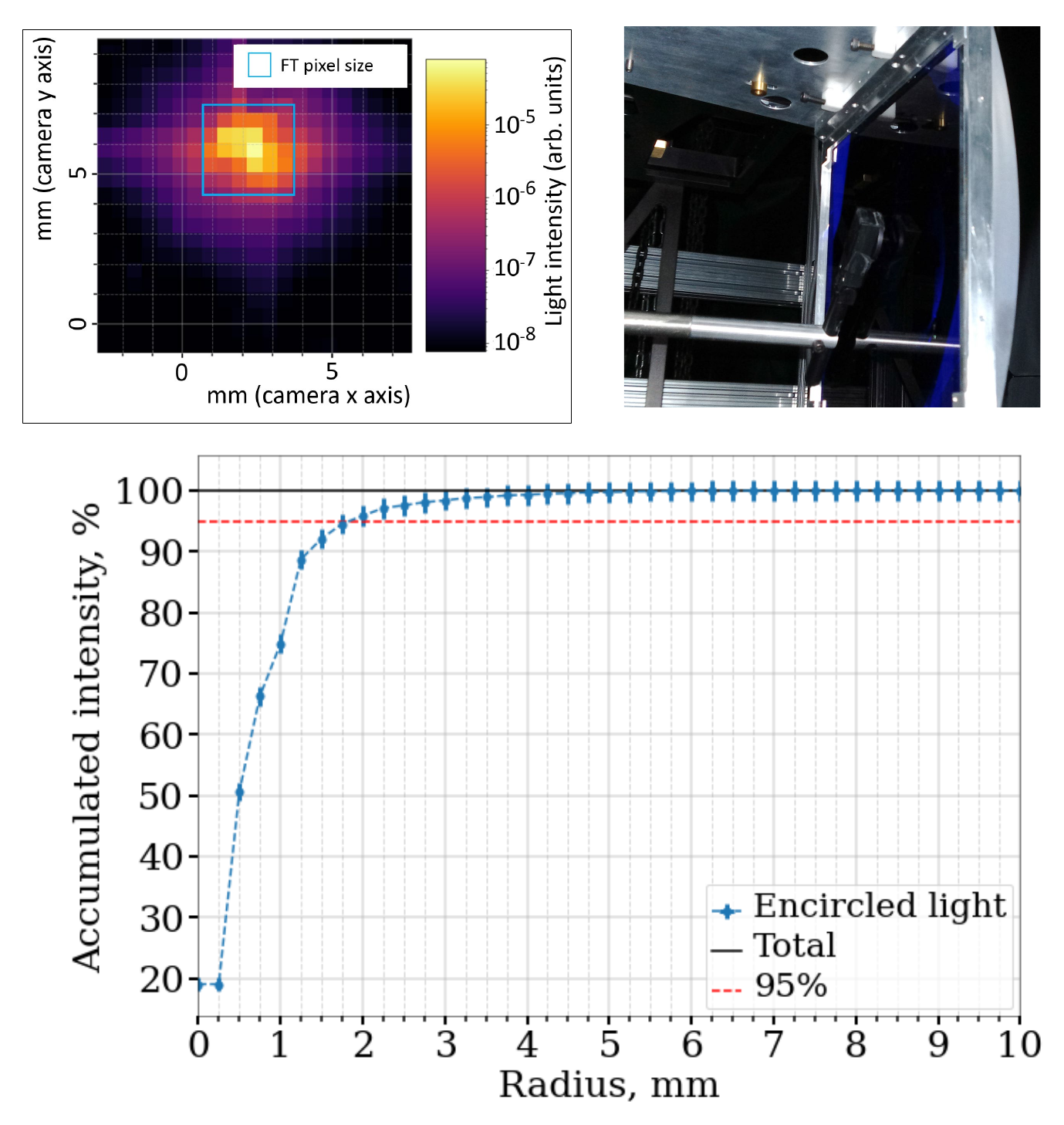}
\caption{Scan of the FT optics PSF using the 1~m laboratory test beam system. Upper Left Panel: Scan map shown on a log scale. Upper Right Panel: The scan was performed just behind the UV filter and camera field-flattening optic (see text). Lower Panel: The fraction of enclosed light as a function of radius. The dashed line at ${95\%}$ is shown for reference.}
\label{fig:FT_psf_scan}
\end{figure}

A PSF scan of the FT optics, including the ACP and all 6 mirror segments mounted and aligned to a common focus, is shown in Figure~\ref{fig:FT_psf_scan}. This scan was made behind a field-flattening optic and BG-3 UV filter that were placed in front of each FT Photo Detector Module (PDM). The BG-3 filter was used to reduce the background outside of the primary UV fluorescence lines of nitrogen in the atmosphere. The field-flattening optic compensated for shape differences between the focal surface which is curved and the front surface of a FT camera module which is flat. More than ${90\%}$ of the PSF can be enclosed within the area of one FT pixel. This measurement was made on-axis at 380~nm. 

A similar measurement was obtained for the CT. The front of the CT camera was smaller than the FT camera and curved in azimuth. No field-flatter optic was required. The wavelength range of the Cherenkov light extended from the UV to the red, and no optical filter was used.  

\begin{table}[htp]
\centering
\begin{tabular}{| l l l|}
\hline
Item & Measurement & Notes \\
\hline
FT PSF & ${1.75\pm{0.05}}$~mm &radius enclosing 95\%\ of light\\
CT PSF & ${1.80\pm{0.05}}$~mm &radius enclosing 95\%\ of light\\
FT Optical Efficiency & ${0.45 \pm 0.01}$ & \\
CT Optical Efficiency & ${0.56 \pm 0.01}$ & w/ BG-3 filter, field flattening optic \\
\hline
\end{tabular}
\caption{Laboratory characterizations of telescope optics using the 1 m test beam at 380~nm. These PSF scans were performed with a common focus.}
\label{tab:optics_tests}
\end{table}

 The alignment of the four mirror CT segments is split into two diagonal pairs. Each diagonal pair focuses parallel light onto a separate spot. The two spots are separated horizontally by $12$\,mm or $0.8^\circ$. This separation is twice the width of the pixel in the CT camera.  Parallel light coming from outside the telescope, for example, the forward-beamed Cherenkov signal of a PeV EAS, will form two spots at the focal surface. In contrast, direct hits on the camera by low-energy cosmic rays or various background fluctuations may initiate a signal on just one pixel. These one-spot vs. two-spot patterns enable the use of a bi-focal trigger algorithm (Section \ref{sec:CT_trigger})  to catch EAS candidates.

%% file: 5_FT.tex
\section{Fluorescence Telescope}\label{FT}
 The FT employs modified Schmidt optics, which are detailed in Section \ref{sec:Optics}. Here we briefly describe the rest of the FT instrument. Further details are available in~\cite{EUSO:SPB2FT:2024a}.

\subsection{FT Camera}

\begin{figure}[h!]\begin{centering}
	\includegraphics[width=0.7\paperwidth]{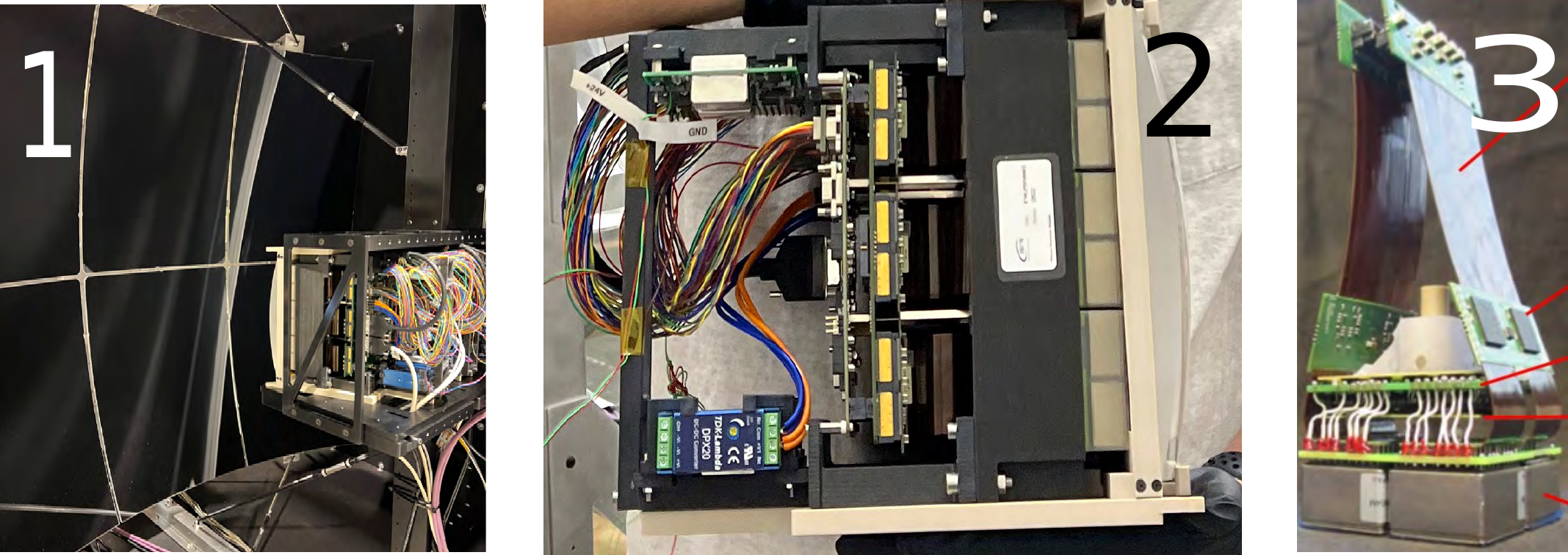}
	\caption{
        View of the inside of the telescope taken during integration (1).
	Assembled photon detection module, behind the BG3 UV filter and field flattening optic (2).
	Assembled EC, prior to potting, showing HVPS control board, ASIC chips and ribbon cable assemblies (3).}
	\label{figs:ft_pics}
\end{centering}
\end{figure}

\begin{figure}[!h]\begin{centering}
\includegraphics[width=0.6\textwidth]{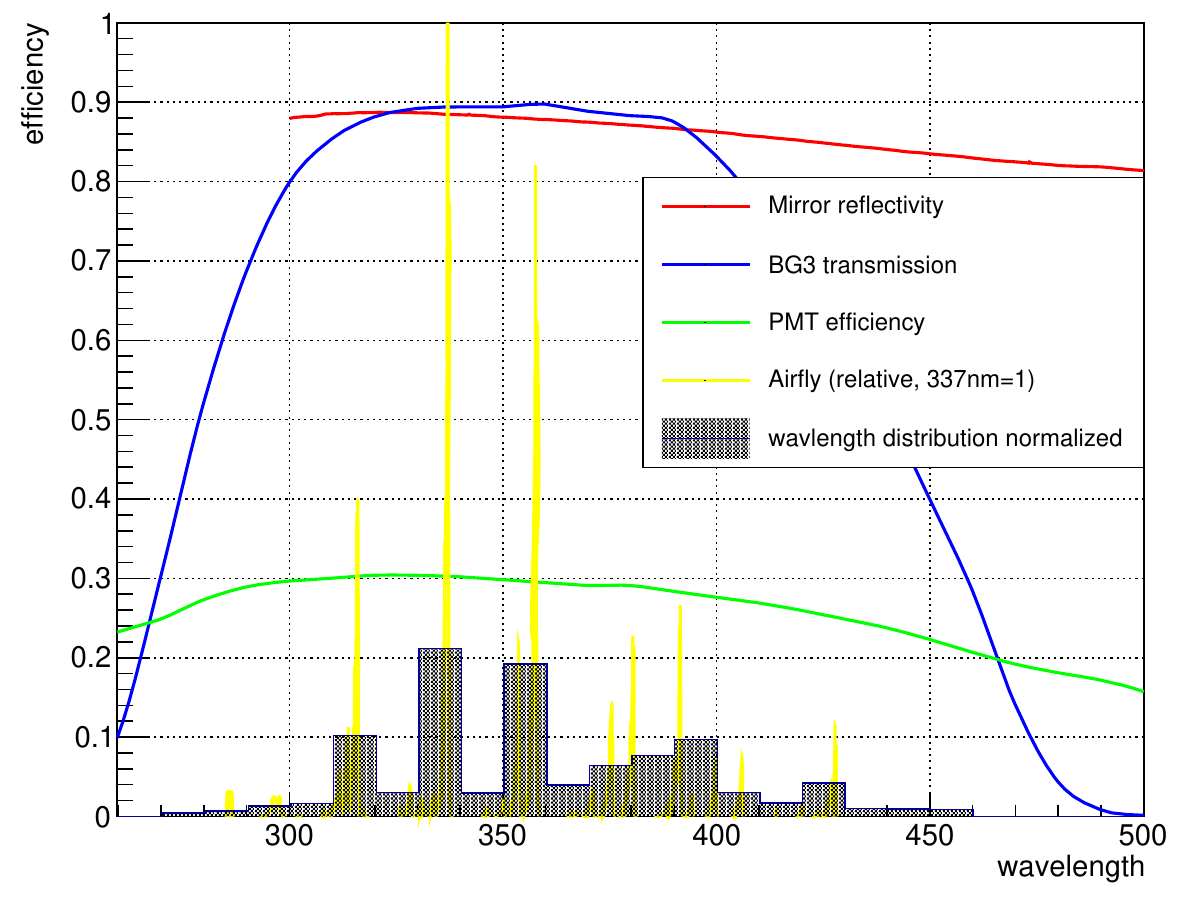}
\caption{Transmittance, reflectivity, and quantum efficiency of the FT system along with the nitrogen fluorescence emission spectrum (arbitrary units) the system was optimized to detect. The total spectrum is normalized to one in the grey histogram. 
The yellow bands show the nitrogen spectrum relative to 337 nm as measured by Airfly~\cite{AIRFLY:2008upr}.}
\label{fig:FT_wavelength}
\end{centering}

\end{figure}

The FT camera system comprises 6912 pixels with individual photoelectron counting capability in 1000 ns time bins. It features three Photo Detector Modules (PDMs) arranged at the focus of the telescope optics system.
The mounted camera and some of its components are shown in Figure \ref{figs:ft_pics}. Key parameters are summarized in Table \ref{tab:spec}. The wavelength dependence of the multi-anode photomultiplier tube (MAPMT) efficiency, BG-3 filter, and optics are plotted in Figure \ref{fig:FT_wavelength}. The nitrogen fluorescence spectrum in air is also shown to illustrate the wavelengths targeted by the FT. These spectral lines are also represented in a normalized histogram of 10 nm-wide bins.

Each PDM is made up of 36 64-channel MAPMTs.  
Four MAPMTs are potted together in a gelatinous compound to form an Elementary Cell (EC) that also contains an integrated high voltage power supply (HVPS). The HVPS employs a Cockroft-Walton circuit. To provide DC coupling, required for photoelectron counting, the MAPMTs are operated with negative HV at the photocathodes. 

Within an EC, the MAPMT outputs are read out and digitized by a SPACIROC3 ASIC \citep{BLIN2018363} capable of counting photoelectron pulses and performing charge integration. The charge integration channel, dubbed KI, allows for measurements of bright signals concentrated in a short pulse beyond the dynamic range of the photoelectron counting channel. The photonelectron count channel is digitized with a 952 kHz frequency and a double pulse resolution of $\sim$10 ns as flown. The KI channels of the MAPMTs are digitized at the same cadence. Eight photoelectron-counting pixels are grouped together to form a KI channel, resulting in 8 KI channels per MAPMT. Data from the 36 ASICs are sent to Artix 7 field programmable gate arrays (FPGA) which multiplexes the data. The PDM is controlled by an Xilinx Zynq\texttrademark 7000 FPGA with an embedded dual-core ARM9 CPU on a custom readout board. This system handles triggering and data packaging, among other tasks. 

\subsection{Data Acquisition System}

\begin{figure*}[bt]
\centerline{\includegraphics[width=0.9\textwidth]{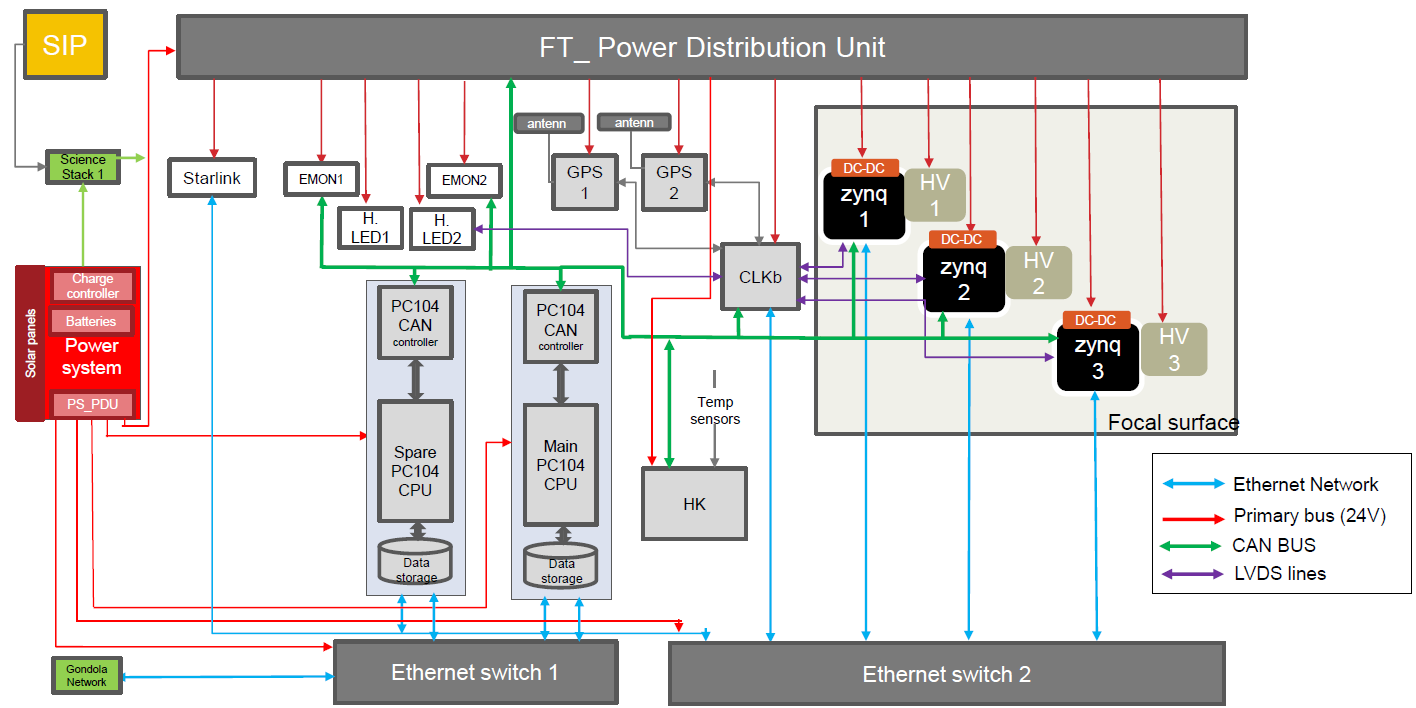}}
\caption{Block diagram of the Data Processing system and its connections with the rest of the instrument for the fluorescence telescope. The camera system with its three photodetection modules are shown in the "focal surface" box on the left. The FT includes two identical CPU systems (left of center) that are used for commanding and data storage. One is active, the other is a spare. The multi-master protocol of the CAN bus facilitates switching between them. A custom clock board (CLKb) synchronizes the GPS-based timing between the 3 PDMs in the camera.
The subsystems critical for turning the telescope on are powered through the gondola power system shown on the left. These include the FT power distribution unit (PDU), the ethernet switches, and the main and spare CPU systems. The CPU controls the FT PDU that powers the rest of the subsystems. The EMON1 and EMON2 modules monitor light levels inside the telescope. The HLED1 and HLED2 modules direct light pulses at the camera for performance monitoring.}
\label{fig:FT_blockdiag}
\end{figure*}

\begin{figure}[bt]
\centerline{\includegraphics[width=0.8\linewidth]{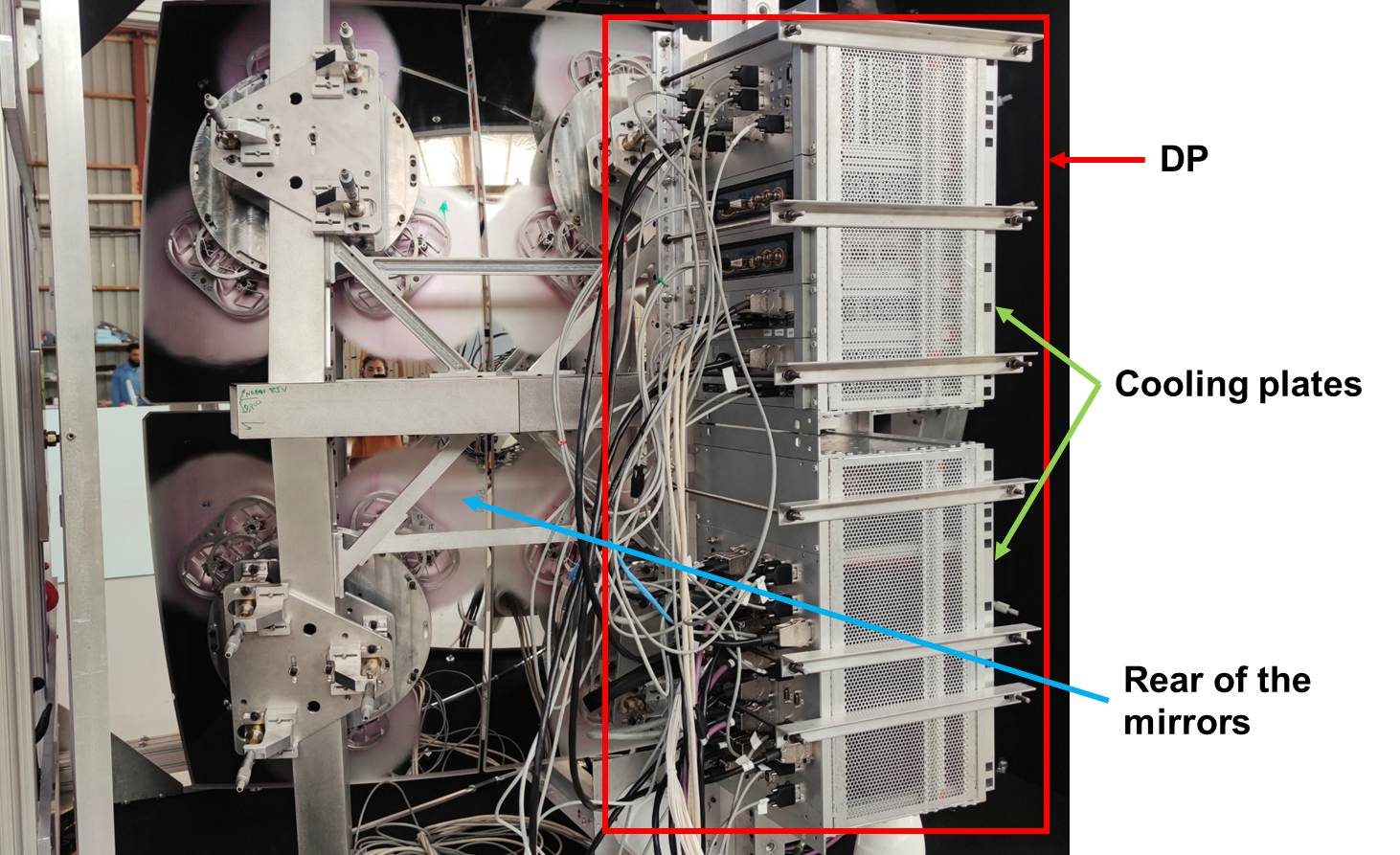}}
\caption{The Data Processing (DP) box, hosting all the Data Processing subsystems, mounted on the back of the telescope structure. The cooling plate is on the back of the box.}
\label{fig:FT_DP_structure}
\end{figure}

The data processing system contains redundant PCI/104 CPUs using Intel\textregistered~Core\textsuperscript{TM}~i7~3517UE cores, 5 SATA disks for data storage, redundant differential GPS systems (Trimble BX992),  a housekeeping board based on an STM32\textsuperscript{TM} microcontroller, and a clock board based on a Xilinx XC7Z System on a Chip. The electronic components were selected to satisfy the requirement of 
being operational over a temperature range of ${+85\degree/-40\degree}$~C or better. The architecture of the full system including ethernet switches and a 24 V DC solid state power controller is shown in Figure~\ref{fig:FT_blockdiag}. Additional details are available in~\cite{Mese_2023}. The protocols of the Controller Area Network (CAN) bus and ethernet connections enabled either CPU to run the system. The multiple network switches facilitated the late addition of the StarLink\texttrademark~ module in W\=anaka. The subsystems and low voltage power supply modules were housed in two custom Eurocard chassis, each equipped with a cooling plate to dissipate heat. This assembly was mounted on the back of the FT mirror system (Figure \ref{fig:FT_DP_structure}). 

The clock board synchronized the readout of the three PDMs by receiving the trigger logic output on each Zynq board and issuing signals to the three PDMs in parallel. GPS data were packaged with trigger and deadtime information for each event that the clock board recorded.  The CPU issued commands to all subsystems, combined the data from the three PDMs and the clock board, compressed these data for downlinking to the ground, and collected FT monitoring data for downlinking. The monitoring data were derived from 18 temperature probes placed around the payload, humidity and pressure sensors, gyroscopes, and from two photodiodes next to the camera \cite{MACKOVJAK2019150}. 

The FT trigger logic and algorithms are described in detail elsewhere \cite{SPB2Trigger}.  The trigger algorithm was designed to search for track-like signatures while recalculating threshold levels every 500 ms. These track-like signatures constituted 2x2 groups of pixels above threshold across 3 consecutive gate time units (GTUs). The trigger algorithm was coded into FPGAs on the PDM Zynq boards. For each global trigger, 128 consecutive GTUs are recorded for each pixel. The first 60 GTUs are before the trigger for pedestal and other studies. The trigger algorithm was tested and refined using simulated data and then tested in the laboratory~\cite{JEM-EUSO:2023nns} and the field using distant laser energy sweeps.

\subsection{Health LED}

Two calibrated health-LEDs (HLEDs) were situated beneath the PDMs to monitor the response of the instrument at four fixed light intensities. The HLEDs functioned as an independent subsystem, flashing a fixed pattern after being activated. This pattern consisted of a flash for 2.5~$\mu$s, a pause for 15~$\mu$s, and another flash at a lower intensity for 5~$\mu$s. The pattern was repeated every 32 seconds. The second HLED was turned on 16 seconds after the first, resulting in a flash pattern from one of the HLEDs every 16 seconds. Aside from providing power, the DAQ system had no connection to the HLEDs. Recording of the HLED signals relied on the internal triggering of the PDMs, making the HLEDs an in situ end-to-end verification of the entire data acquisition (DAQ) system. 

\subsection{Fluorescence Telescope Field Campaign}
\begin{figure}[!ht]
\includegraphics[width=1.0\textwidth]{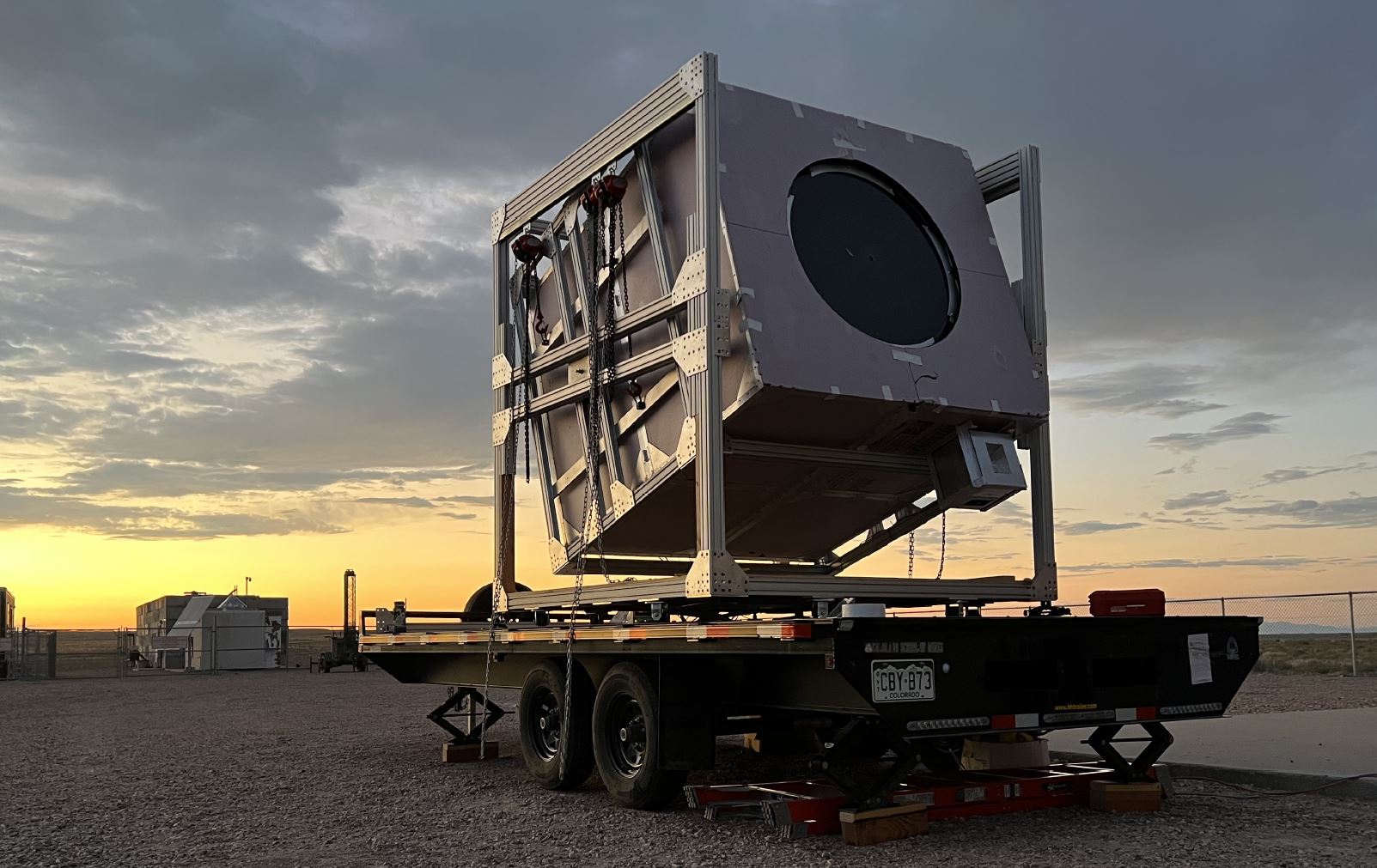}
\caption{The FT in the Utah desert for field tests. Photo: George Filippatos}
\label{fig:FT_field_photo}
\end{figure}

Prior to payload integration in Colorado, the FT was field tested at the Telescope Array Black Rock Mesa site in the West Utah Desert over a 12-day campaign. The night sky background, including stars, meteors, airglow, and airplane strobe lights, was observed for tens of hours. 

The EUSO-SPB2 FT was cradled in a specially designed ``yoke" frame and transported on a flatbed trailer between Colorado and Utah. The FT cradle could be manually rotated on twin bearing assemblies to an elevation angle ranging from about 10\degree~below horizontal to vertical. An elevation pointing accuracy of about 0.1\degree~could be achieved using digital carpenter levels. The yoke could be manually rotated in azimuth about a vertical pivot on the trailer to achieve full-sky pointing of the FT. The field test configuration is shown in Figure \ref{fig:FT_field_photo}. The mirror segment assemblies were dismounted in Colorado, crated for transport, and remounted at the field site. 
 
A portable field laser system housed in a trailer was taken to Utah for the campaign. Optical UV track-like signatures mimicking EAS were generated by light scattered out of the beam of the field laser system, which was shot into the sky at various angles. The laser was a frequency tripled Nd-YAG operating at 355~nm with a pulse duration of approximately 10~ns, and an adjustable energy of 0.1 to 70 mJ/pulse. The system included a robust automated beam-steering periscope, as well as beam energy monitoring and calibration subsystems. Firing times were synchronized to a GPS clock \citep{Hunt:2016do}.  The same laser system and site had been used in 2016 for EUSO-SPB1 FT field tests~\cite{Adams_2021}.

\begin{figure}
\centering
    \includegraphics[width=0.8\textwidth]{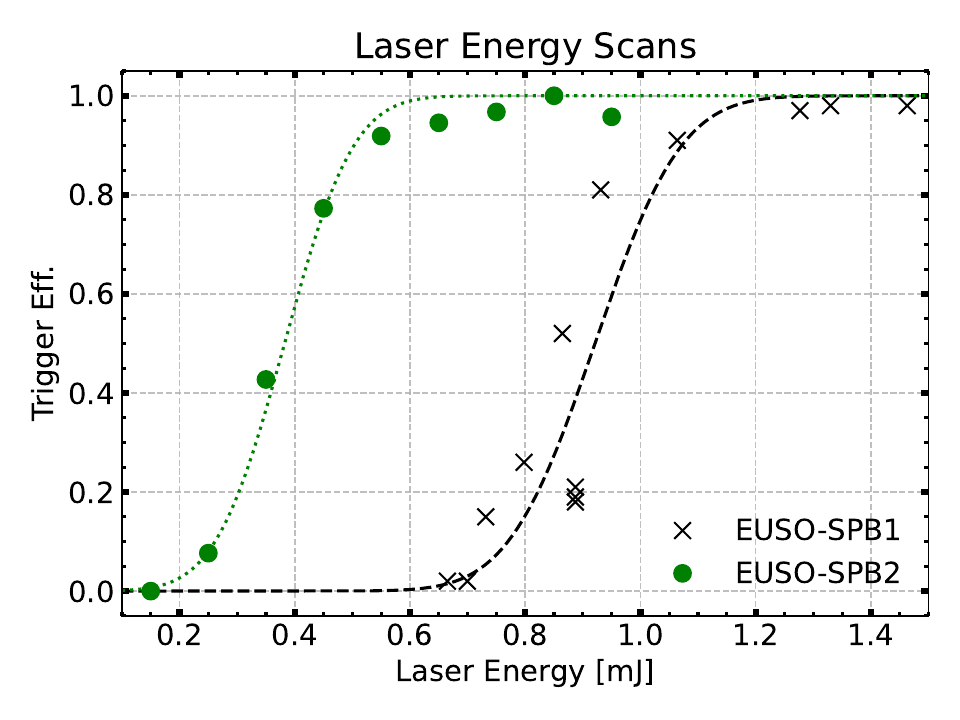}
    \caption{Energy threshold measured with a laser during the field campaigns of EUSO-SPB2 (green) and EUSO-SPB1 (black) \citep{Adams_2021}. The data are fitted with two parameter fits of the form $T(E)=\frac{1}{2}\big[1+\text{erf}[(E-A)/B]\big]$. Each data point represents the number of laser pulses that triggered the FT divided by the 100 laser pulses that were fired. }
    \label{fig:laserEff}
\end{figure}

A benchmark measurement of the FT trigger threshold, one of the most important field tests, was conducted using the laser system.
This benchmark test is detailed in~\cite{EUSO:SPB2FT:2024a}. 
The energy at which the laser began to trigger the FT was established by firing batches of 100 shots at ever-decreasing energies in a constant geometrical configuration.
The laser was positioned 24~km from the detector and the beam was aimed 45\degree~above the horizon and away from the FT. The FT was pointed 10\degree~above the laser trailer location. The distance and angle of the laser beam in the FT FOV is similar to the typical distances and angles of EASs that could be observed during flight.
The fraction of laser shots that trigger the FT at each energy is shown in Figure \ref{fig:laserEff}, together with the energy scan conducted in 2016 for EUSO-SPB1, which used the same laser at the same distance and orientation. This test established a threshold energy for the EUSO-SPB2 FT that was a factor of two below the same scan of the EUSO-SPB1 FT.   

\begin{figure}
  \centering
    \includegraphics[width=1.0\textwidth]{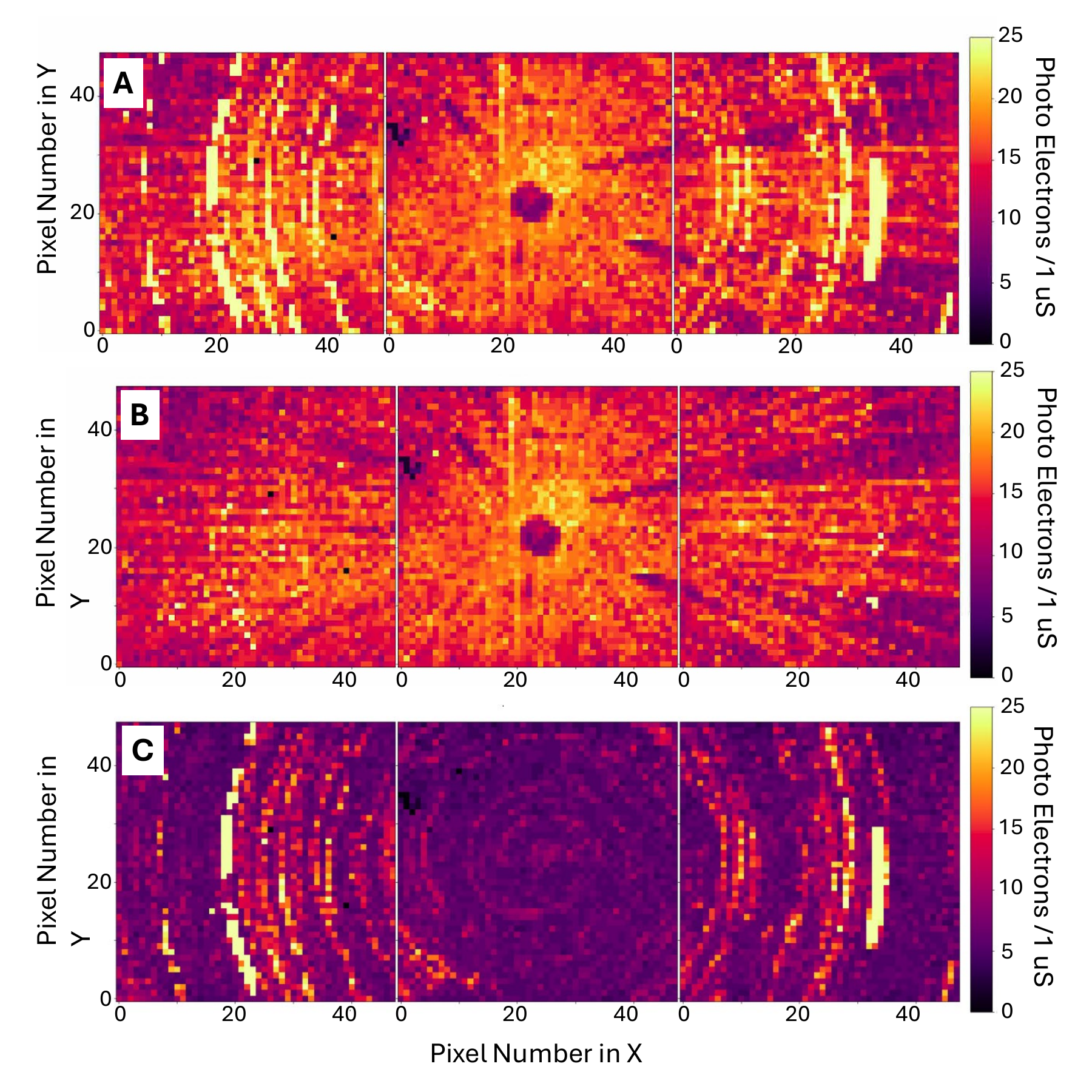}
    \caption{Examples of laser tracks and stars recorded with the FT pointing vertically and rotated 360\degree~ in azimuth (see text).}
    \label{fig:FT_rotate_field_test}
\end{figure}

Other laser scans were performed to explore the behavior of the FT across its field of view to natural and artificial light sources. In one example, the laser beam was aimed almost directly over the FT at an elevation angle of ${10\degree}$ from 24 km and pulsed continuously. The FT was pointed upward and rotated in azimuth below the laser beam path. Over a 360\degree~rotation, 2095 triggered events were recorded. Three representations of the patterns seen in the camera pixels are shown in Figure \ref{fig:FT_rotate_field_test}. The brightest value of each pixel over the 128 time bins ${\times}$ 2095 triggers, displayed in panel A, shows the straight lines of laser shots, an offset straight track near the camera center, likely caused by a meteor, and circular patterns from bright stars. A procedure was developed to separate the straight tracks from the star patterns. In panel B most of the star signatures are removed. Pixels for which the photoelectron count remained above some threshold for all 128 GTUs of an event were tagged as ``star-struck" for that event. The brightest pixel values displayed in panel B are selected from events when the pixel was not tagged as star-struck. The threshold used was 1 photoelectron count. In panel C, to highlight bright star paths, the brightest pixel is selected from the first 30 time bins that were recorded before the trigger. At a minimum, this study found no obvious problems in the end-to-end pixel mapping or in the triggering across the camera, and found a way to tag "star-struck" pixels.

Calibrated pulsed LEDs placed both in the near and far fields were used to provide an alternative calibration to the piecewise method used in the laboratory.  An in-depth discussion of the field test and the FT calibration is presented in~\cite{Kungel:2023C6}.

%% file: 6_CT.tex
\section{Cherenkov Telescope}\label{sec:CT}   

\begin{figure}[!hb]
    \includegraphics[width=1.0\textwidth]{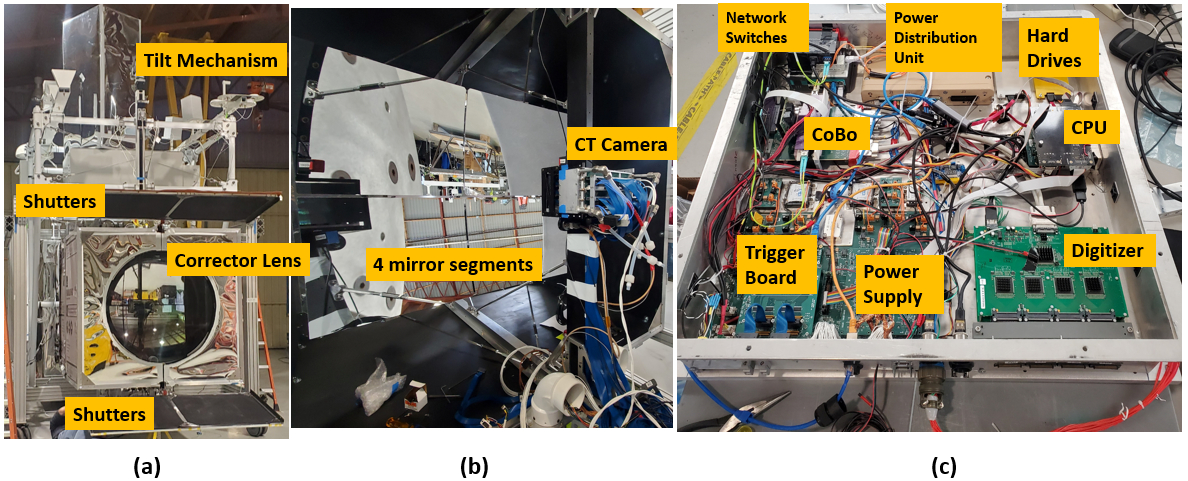}
    \caption{The CT during preparation for launch in W\=anaka, New Zealand. Panel (a): CT is shown shutters open as attached to the gondola. Panel (b): The inside of the telescope is shown with the CT camera mounted the focus of the 4 mirror segments. Panel (c): The control, digitization, and triggering hardware in an enclosure that attached to the bottom of the telescope for flight.}
    \label{fig:assembled_components}
\end{figure}

\begin{figure}[!ht]
\centering
\includegraphics[width=1\textwidth]{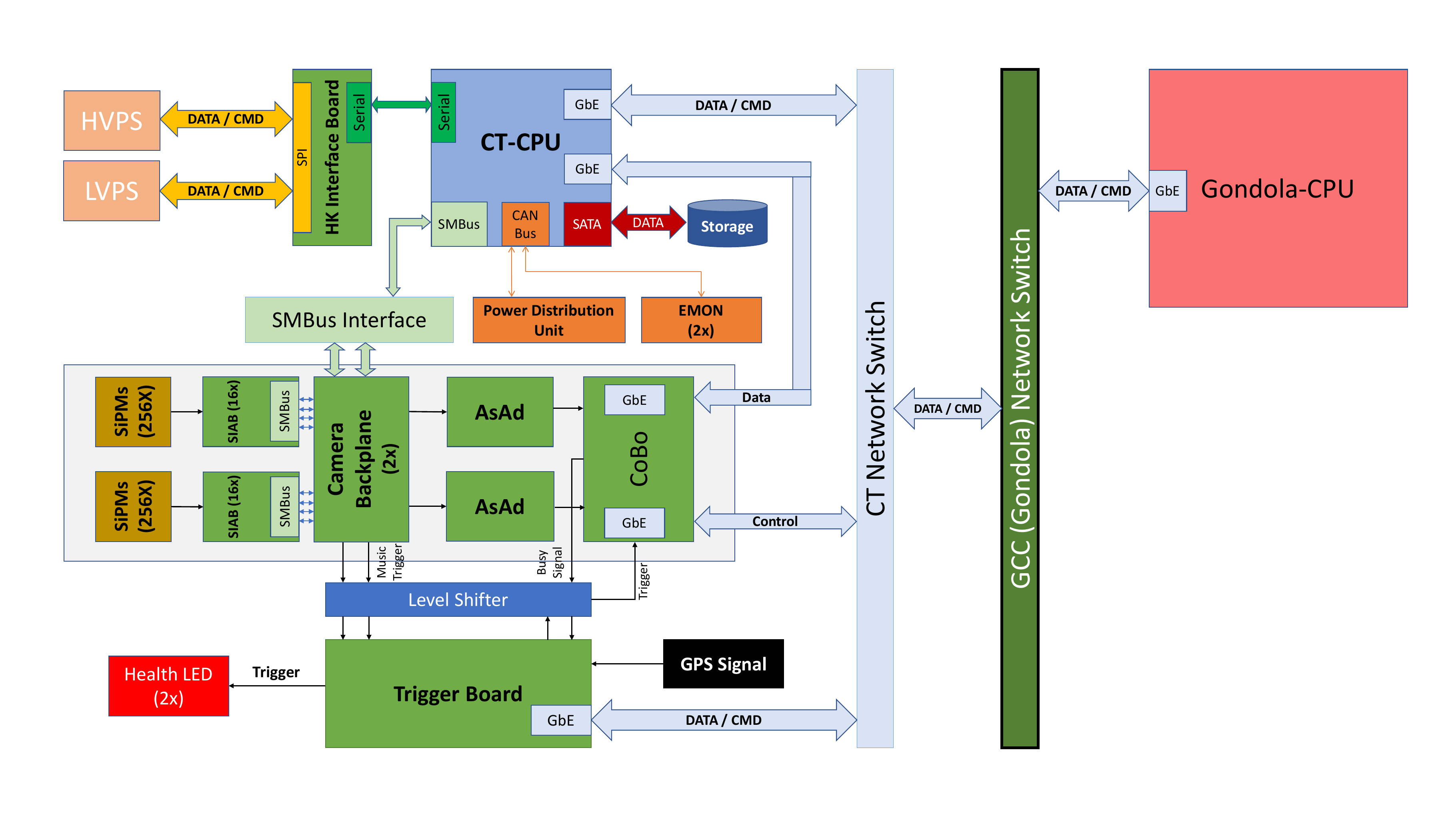}
\caption{System architecture of Cherenkov telescope camera and data acquisition. The subsystems grouped in the shaded rectangular include the SiPMs that convert light to electrical signals. These signals are then shaped and amplified by custom 8-channel (SIAB) boards that also generates a fast trigger as an "OR" of the 8 channels. The fast "OR" signals are routed through the camera backplane to the trigger board. The 512 analog signals are sent to custom Support \& Analog-Digital conversion circuit boards (AsAd)  for digitization and triggering. The digitization is performed by switched capacitor arrays with a 100 Mhz sampling rate. The data Concentrator Board (CoBo) collects the traces, adds time-stamping and performs zero suppression. The entire system is controlled by a single board computer (CT-CPU) which is the same model used in the FT.}
\label{fig:CT_architecture}
\end{figure}

The CT was optimized to detect fast flashes of forward-going Cherenkov light from PeV energy EASs. Key parameters are listed in Table~\ref{tab:spec}. The optics, including the bi-focal focusing, are described in section \ref{sec:Optics}. The CT and its components are shown in Figure \ref{fig:assembled_components}. The hardware architecture is shown in Figure~\ref{fig:CT_architecture}.
\vspace{0.1cm}

\vspace{0.1cm}

\subsection{Camera}

\begin{figure}[!ht]
\includegraphics[width=1.0\textwidth]{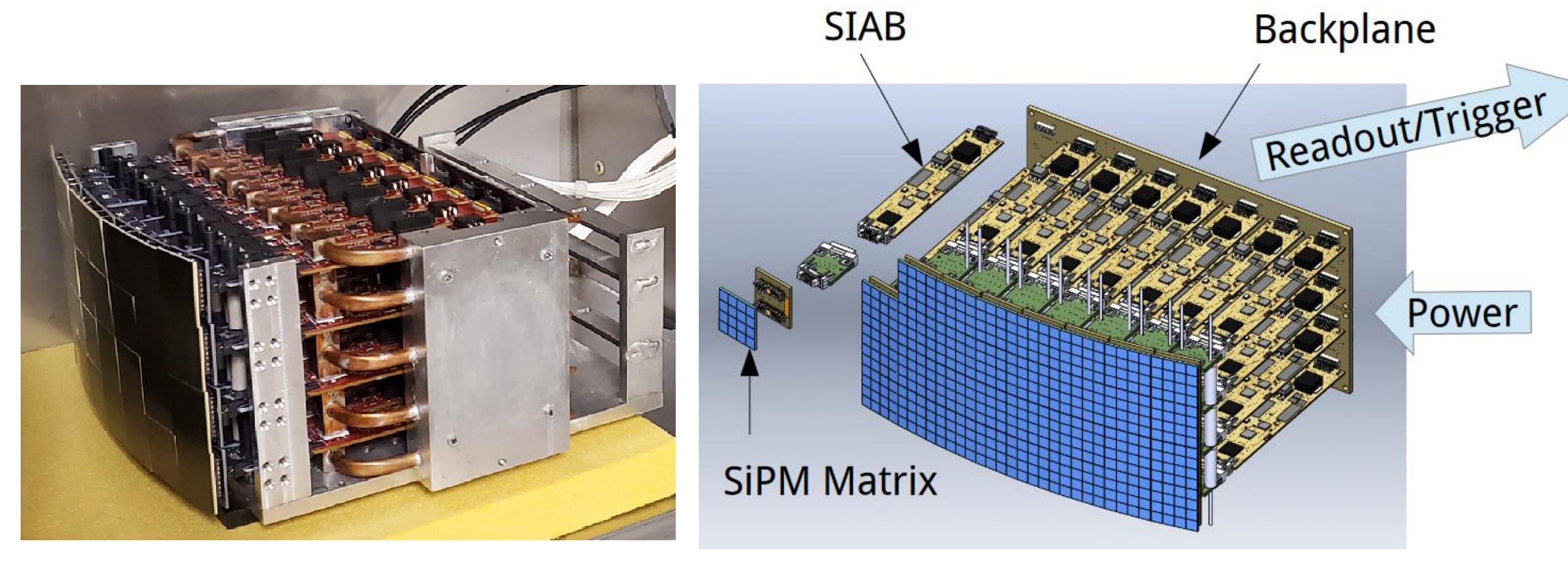}
\caption{The CT camera features 4x4 SiPM matrices that are tiled to match the azimuth curvature of the focal surface with 16$\times$32 pixels.}
\label{fig:CT_camera_rend}
\end{figure}

\begin{figure}
    \centering
     \includegraphics[width=0.7\textwidth]{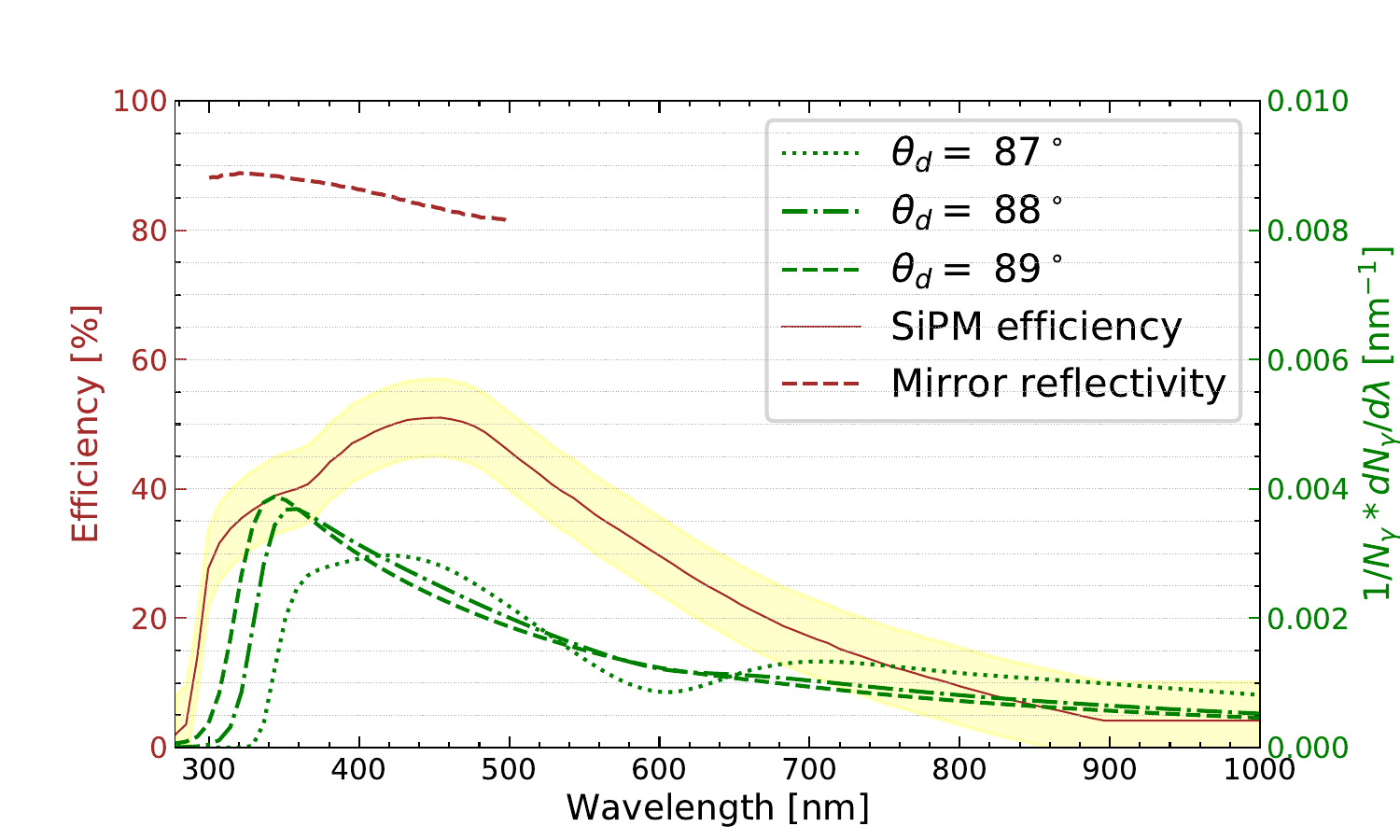}
    \caption{Quantum Efficiency of the Hamamatsu S14521 SiPM, adapted from manufacturer data sheet and the spectral distribution of Cherenkov light from 10 PeV EASs initiated by cosmic rays at 80g/cm$^{2}$ starting grammage, for different angles, $\theta_{d}$ from nadir, after atmospheric attenuation is taken into account.} 
    \label{fig:CT_Spectral_Response}
    \end{figure}

The CT camera (Figure \ref{fig:CT_camera_rend}) has 512 SiPM pixels arranged in a 16${\times}$32 grid to cover a $6.4^\circ \times 12.8^\circ$~(vertical and horizontal) FoV. Hammatsu S14521-6050AN-04 SiPM units were selected. They come packaged as 16 $6 \times 6 \rm{mm}^2$ pixels arranged in a $4\times4$ matrix. The camera has 32 SiPM matrix units. The Photon Detection Efficiency (PDE) and the EAS wavelength dependence are illustrated in Figure \ref{fig:CT_Spectral_Response}. Since the Cherenkov light signals from EASs extend into the visible range, no optical filter is used. As the camera modules are thin and relatively small they can be positioned to approximate the azimuth curvature of the focal surface without introducing large gaps. The small defocusing effect on the PSF is readily handled by the larger dimensions of the CT SiPM pixels compared to the FT MAPMT pixels. No field flattening optics was required for the CT camera.  For a detailed description of the CT camera system, including laboratory characterizations and calibrations, see \cite{CT:technical:2024a}.

Each SiPM matrix unit is connected to a custom-designed Sensor Interface and Amplifier Board (SIAB) populated with two Multipurpose Integrated Circuit (eMUSIC) chips \cite{GomezMUSIC:Arrays}. The eMUSIC chips shape and amplify the SiPM signals which are then guided to a custom backplane. 

\begin{figure}[!ht]
   \centering
\includegraphics[width=0.8\textwidth]{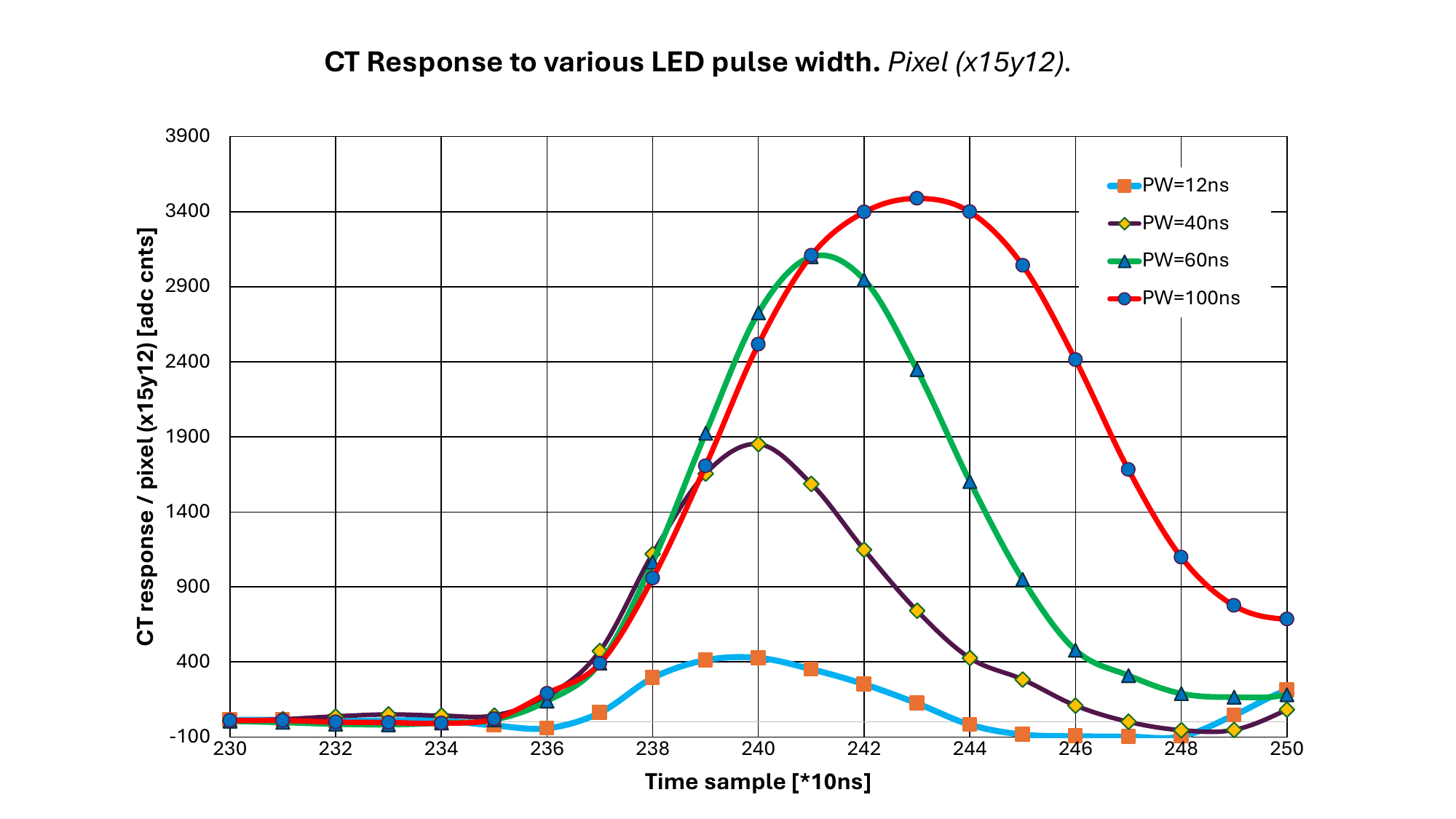}
\caption{Example response of a CT pixel to an external LED for different pulse widths (PW) in a outdoor preflight test at W\=anaka.}
\label{fig:CT_LED_JimK_Wanaka}
\end{figure}

A sample of a digitized signal collected pre-flight in W\=anaka using an external pulsed LED that illuminated the CT entrance pupil is shown in Figure \ref{fig:CT_LED_JimK_Wanaka} for different LED pulse lengths. The effect of the signal stretching in the short pulses can be seen. 

\vspace{0.1cm}

\subsection{Triggering} \label{sec:CT_trigger}
The dual-spot focusing of parallel light from outside the telescope supports the use of a bi-focal trigger to discriminate against accidental triggers. The pulses from EASs in the bifocal spots are correlated in time in contrast to fluctuations in the night sky background. (For reference, the dark count rate of the SiPMs is very low relative to the NSB rate.)

There are two steps to the CT trigger system: the discriminator and the trigger board. The discriminator used is inside each of the MUSIC chips. It is a ``Fast OR circuit" which includes each of the 8 pixels connected to the MUSIC. In the case of EUSO-SPB2, these 8 pixels correspond to a 2x4 column in each of the matrices. When any of these 8 pixels goes over the configurable threshold in the MUSIC discriminator, a square pulse is created and stretched over the time the pixel is over threshold. The square pulse is then fed into a level shifter board, which adjusts the signal level to Transistor-Transistor Logic (TTL) level and then feeds this signal to the trigger board, which contains the trigger logic.

The trigger connects each of the discriminators to an individual channel. When a trigger from the discriminator is received, only the rising edge is used and the trigger signal is stretched to 50~ns. If a trigger is also received from a laterally adjacent MUSIC within this 50~ns interval, then a readout trigger is issued, and the event is saved to disk.

\subsection{Data Acquisition}

Cables carry the analog signals from the camera to digitizer boards that are located inside the Main Electronics Box (MEB), which is mounted below the telescope and shown in Figure \ref{fig:assembled_components}. The digitization is performed by ASICs developed for the General Electronics for Time projection chambers (GET)~\cite{Pollacco:2018bsd} project. The AGET samples the signals with a rate of 100MSa/s and a buffer depth of 5.12$\mu s$. Upon receiving a readout command from the trigger board, the signals are digitized with 12-bit resolution. The digitized signals are managed by a Concentration Board (CoBo). The CoBo applies time stamps, performs zero suppression, and compresses the digitized signal. The CT camera and readout are controlled through the CT-CPU located inside the MEB, which also stores the data before being downloaded. 

\subsection{Shutters and Tilting}
A pair of motorized mechanical shutter assemblies was developed to protect the CT camera from direct sunlight. The shutters were mounted in front of the CT entrance pupil. A tilting mechanism used a linear stage to raise and lower the front of the CT about its pivot attachment at the back of the telescope. Both systems used stepper motors and limit switches interfaced to a Siemens Programmable Logic (PLC) Controller. The PLC controlled three custom driver boards that send current steps to turn the motors. The PLC automatically closed both shutters after a time interval that was specified in the ``open shutters" command. Two tilt sensors mounted on the top of the CT C frame structure measured the CT tilt angle.

\subsection{Calibration}
 The bench testing and calibration of the CT camera and readout are detailed in \cite{CT:technical:2024a}. The calibration includes measurements of the PDE at different bias voltages and angles of incidence for wavelengths from 200\,nm to 1000\,nm, as well as a measurement of the linearity and dynamic range of the analog signal chain. The CT also included a pulsed HLED system to monitor the relative performance of the camera.

\subsection{Cherenkov Telescope Field Campaign}
The CT was transported to the Black Rock Mesa site of the Telescope Array experiment near Delta Utah at the end of February 2022 and reassembled for a week of field tests. Following a period of on-site commissioning and mirror alignment, the CT recorded near-field pulsed LED events, overhead laser track events, and TeV-scale EAS events.

A simple method to align the mirror segments was implemented in situ by positioning a pickup-truck about 4.5~km from the telescope. The telescope was rotated in its ground support frame on the trailer so that the light from one headlight was focused at the center of the focal surface. The other headlight was covered. At this distance the 8.5~cm radius of the headlight yields an angular spread of light across the CT entrance pupil that is about 25 times smaller than the FoV of one CT pixel. The 4 CT mirror segments were then aligned to the desired bifocal configuration of two spots separated horizontally by 12~mm using a paper target on the focal surface. The light deployed for this method is bright enough to be used at dusk as well as at night. 

\begin{figure}[!ht]
\includegraphics[width=1.0\textwidth]{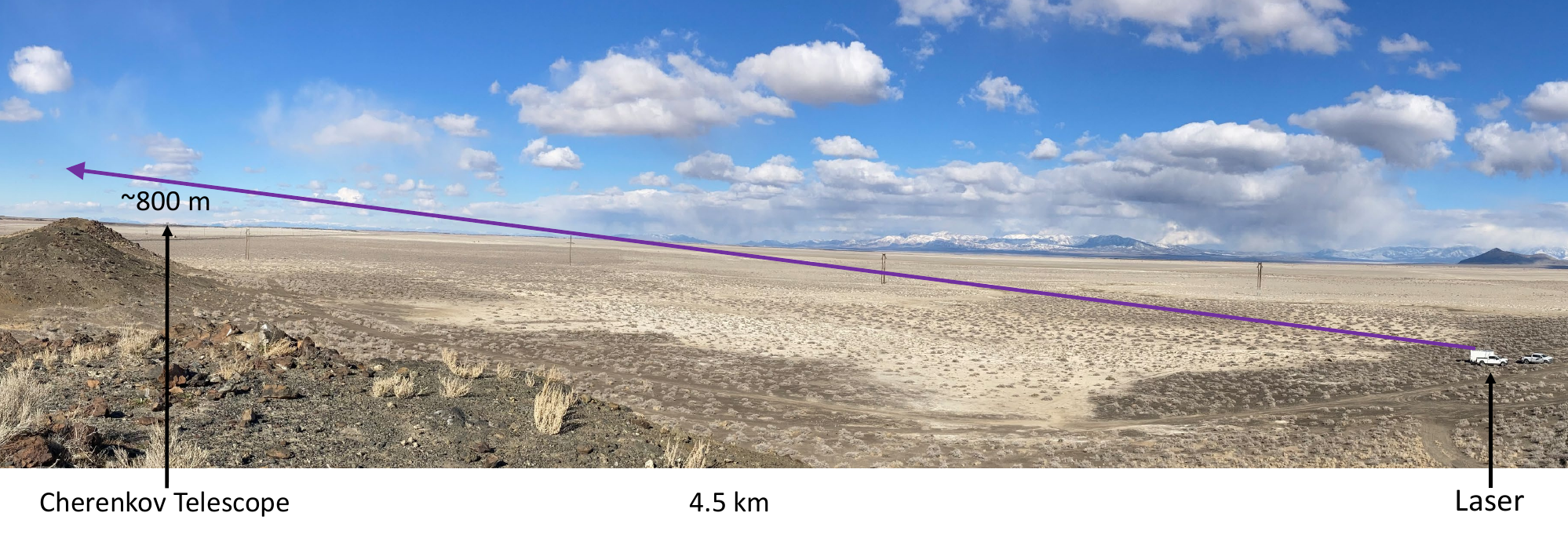}
\caption{The position of a laser system and the CT for tests of track-like events. (Photo: L. Wiencke)}
\label{fig:CT_field_laser_photo}
\end{figure}

\begin{figure}[!ht]
\centering
\includegraphics[width=0.8\textwidth]{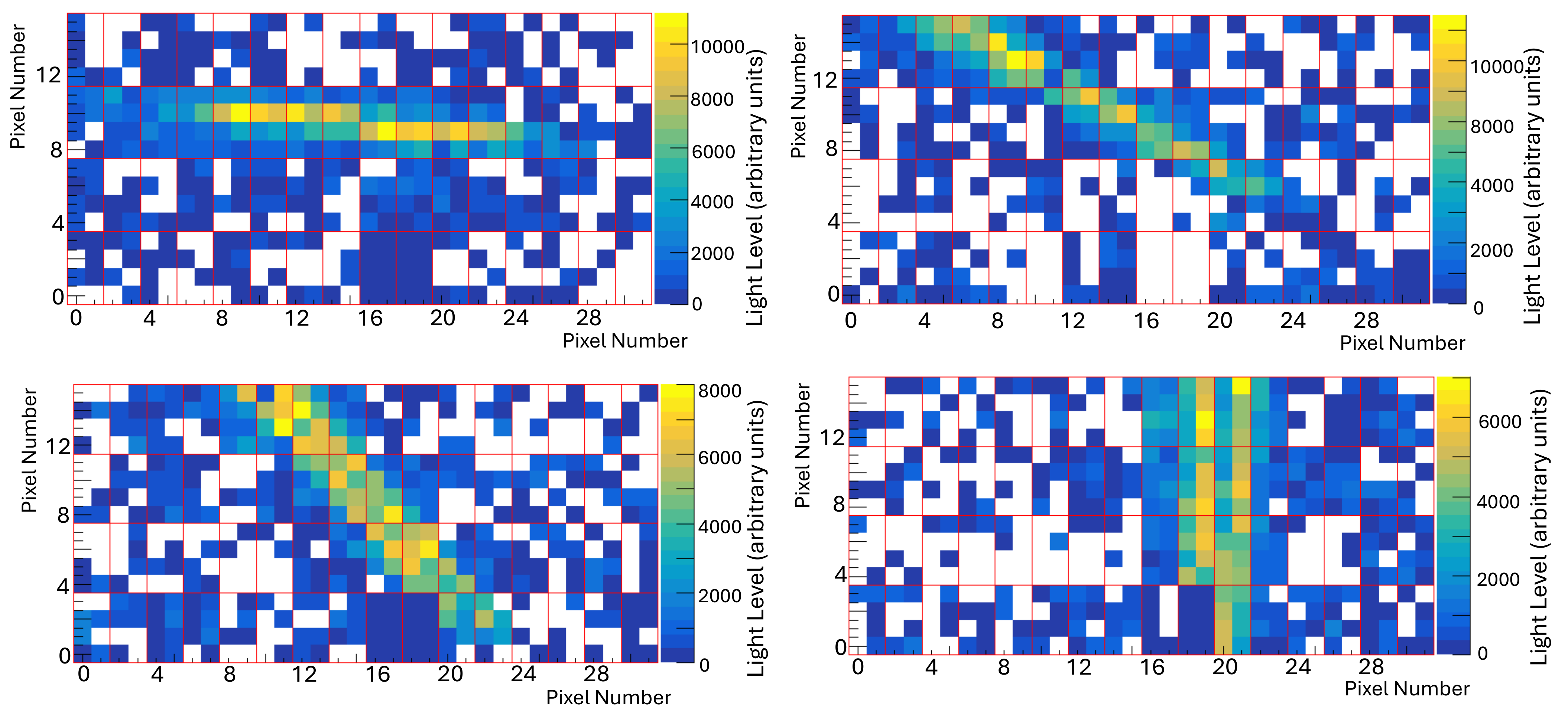}
\caption{Examples of laser tracks recorded by the CT in the field.  The CT was pointed up and rotated in azimuth under the fixed-direction laser beam. The track pattern shows the widest separation when the CT bifocal alignment is perpendicular to the laser beam (lower right panel).}
\label{fig:CT_field_laser_tracks}
\end{figure}

A test of the camera and the bifocal alignment used laser tracks fired over the telescope with the telescope pointed vertically (Figure~\ref{fig:CT_field_laser_photo}). For this test, the laser was triggered on the GPS second. The camera was triggered externally with a GPS-based trigger. A timing offset was added to compensate for the light travel time so that the 10~ns laser pulse crossed the CT FOV near the middle of the CT 5~$\mu$s readout window. Examples of laser tracks in Figure~\ref{fig:CT_field_laser_tracks} for different telescope azimuth angles illustrate how the bifocal separation is largest when the 12 mm spot separation is perpendicular to the laser direction and decreases as the parallel orientation is reached. To the best of our knowledge, this is the first time that any Cherenkov Telescope has been tested with overhead laser tracks. 

\begin{figure}[!ht]
\centering
\includegraphics[width=0.8\textwidth]{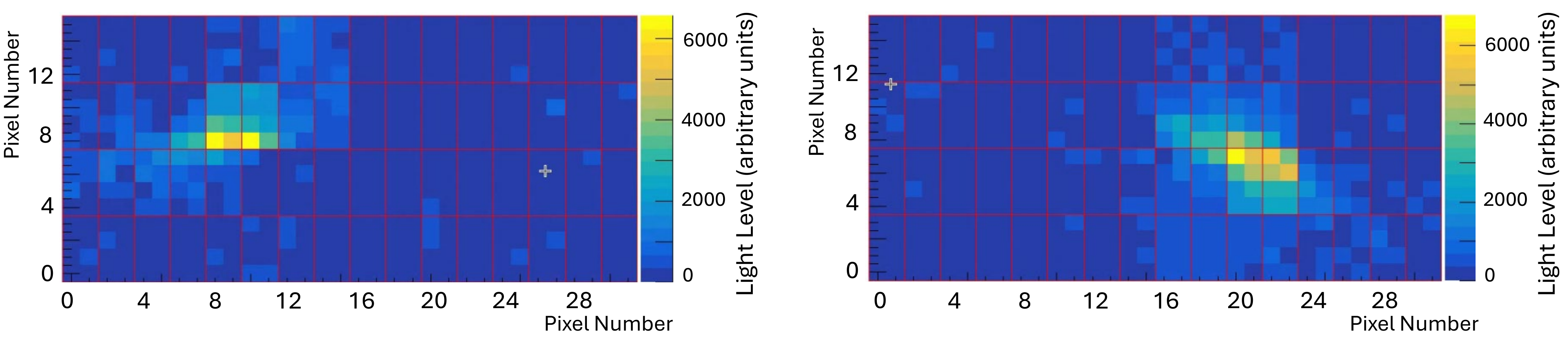}
\caption{Examples of extensive air showers record with the CT during field tests at the Telescope Array site near Delta Utah, USA.}
\label{fig:CT_field_cosmics}
\end{figure}

With the telescope commissioned, including the bi-focal trigger, and pointing up, the first bi-focal signals easily recognized above background (Fig. \ref{fig:CT_field_cosmics}). These signals disappeared when an opaque cover was temporarily placed over the entrance pupil and then reappeared after the cover was removed.

The recorded rate of these events was a few per minute, which is consistent with the expected rate of downward-going EASs in the 10 TeV range. This estimate folds in the known cosmic ray flux, the CT FoV, the photon collection efficiency of the CT optics with the 4 mirror segments aligned bi-focal mode, the optical efficiency and camera efficiency, and the diameter of the ground-level optical footprints of simulated air showers. The CORSIKA 7 simulation package was configured for photon primaries at 1, 10, and 100 TeV energies at the 1.4 km altitude of Delta Utah. The 1 TeV showers are too dim. The expected rate of EASs above 100 TeV is about 0.1/min. The corresponding number for 10 TeV EASs is 8/min, which is about 4 above the observed rate. We note that interpolating to 30 TeV yields an expected rate of 2/min.

This energy range is much lower than the PeV range (Table \ref{tab:spec}) predicted for flight. This is because the Cherenkov light observed looking up from the ground is produced much closer (~10 km) to the detector. Its footprint on the ground extends for a few 100 m. The Cherenkov light observed with the CT pointing horizontally during flight is produced 100s of km away, experiences larger atmospheric losses, and extends across several km transversely at the detector. 

%% file: 7_UCIRC.tex
\section{Infrared Camera}\label{sec:Infrared-Camera}
The presence of high clouds within the FT FoV can reduce the UHECR event detection rate because such clouds can block some or all of an EAS optical signal. Determining the exposure of the FT accurately to UHECRs thus requires knowledge of the volume of atmosphere within the FoV, above clouds.  

\begin{figure}
    \centering
    \includegraphics[width=0.8\textwidth]{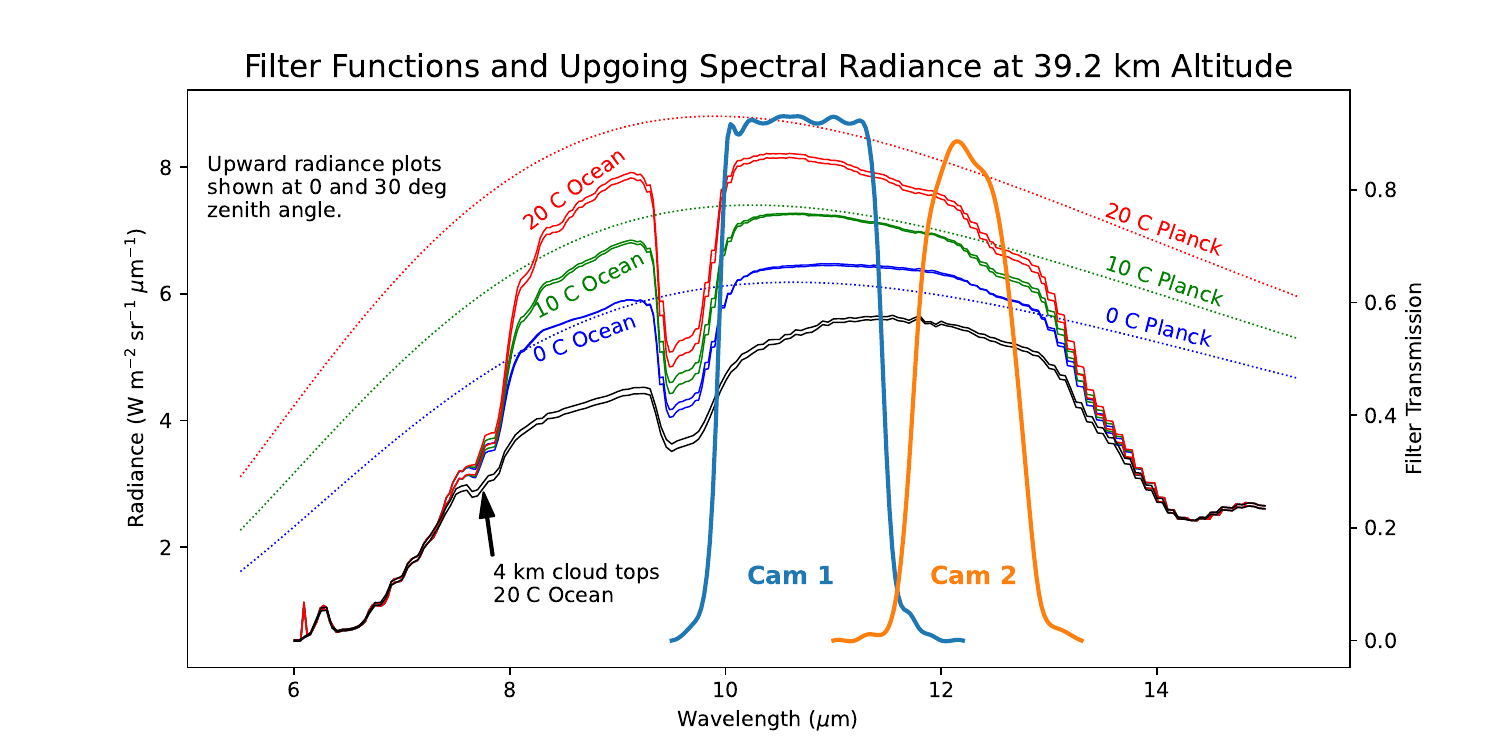}
    \caption{Upgoing spectral radiance as a function of wavelength as calculated using COART \cite{jin+06}, assuming different ocean temperatures with no clouds (red, green, and blue lines) and the presence of clouds with tops 4 km above a 20 C ocean (black lines). The two solid lines shown in each case correspond to 0 and 30 degree zenith angles. The blackbody curves corresponding to each ocean temperature are also shown for reference (dotted lines), as are the bandpasses of UCIRC2's two filters.}
    \label{fig:COART}
\end{figure}

UCIRC2 pointed in the nadir direction to collect information about the cloud coverage and altitude (Cloud Top Height, CTH) within the field of view of the FT. Because the clouds are at the temperature of the air, CTH can be inferred from the cloud temperature $T_{\rm c}$, which can be estimated using two brightness temperatures in bands near the wavelength of the cloud blackbody peak. A calibrated image in a single frequency band can be used to determine the temperature of an object of known emissivity ($\epsilon$), but the cloud emissivity is highly variable and significantly less than 1. Thus, a multifrequency observation is required to break the degeneracy between $\epsilon$ and $T_{\rm c}$. For a single layer of clouds above an ocean of known surface temperature and reflectivity (and thus power, $P_{\rm E}$), one can estimate the power on the detector, $P_{\rm tot}$ as $P_{\rm tot} = \epsilon P_{c}+(1-\epsilon)P_{E}$ where $P_{c}$ is the power of the cloud, from which $T_{c}$ and thus CTH can be inferred.  For a more precise calculation, the Coupled Ocean-Atmosphere Radiative Transfer Model (COART) presented in \cite{jin+06} provides the radiance at any frequency by solving the radiative transfer equation from the ocean to a specified level in the atmosphere, including clouds of arbitrary altitude and emissivity (see Figure \ref{fig:COART}). Other methods for reconstructing CTH can be found in \cite{anzalone+19}.

UCIRC2 is equipped with two $640\times480$ pixel uncooled IR cameras with 14mm lenses, focused at infinity. The cameras have a $42^\circ \times 32^\circ$ FoV, chosen to be somewhat larger than the FT FOV. One camera is fitted with a 9.6 to 11.6$\mu$m bandpass filter and the other is fitted with a 11.5 and 12.9$\mu$m bandpass filter. More details can be found in~\cite{Diesing:2020exu}.

%% file: 9_Flight.tex
\section{Flight}
\label{sec:Flight}

\subsection{Operational Plan}
\begin{figure}[!h] \begin{centering}
\includegraphics[width=0.8\textwidth]{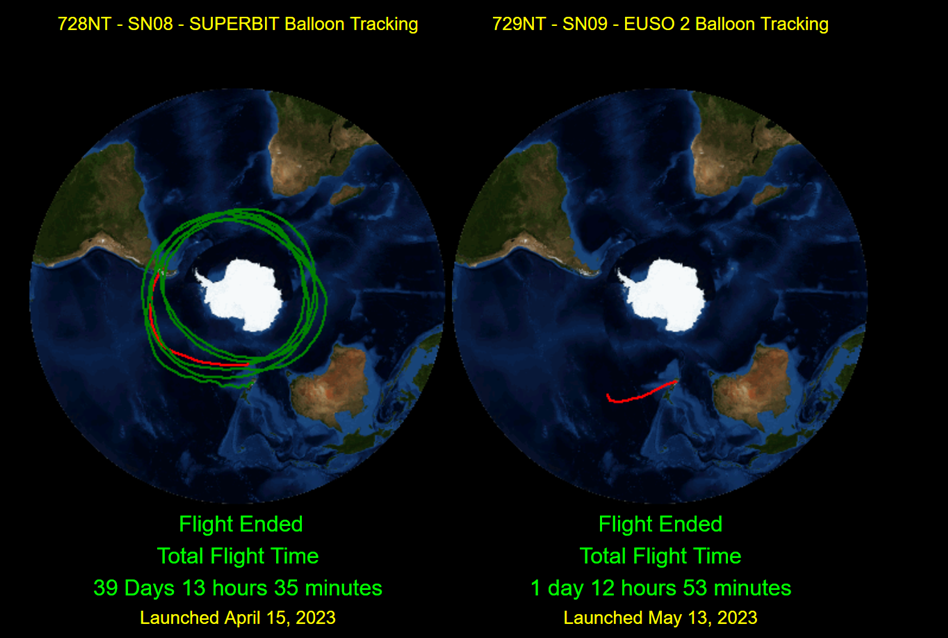}
\caption{Two 18 MCF super pressure balloons were launched from W\=anaka NZ during the 2023 campaign. The trajectory of the EUSO-SPB2 flight is shown on the right.}
\label{fig:Trajectory}
\end{centering}
\end{figure}

The operations were planned for a long-duration flight of more than 50 days to observe UHECRs with the FT, to observe High Energy Cosmic Rays (HECRs, 1 PeV to 1 EeV) with the CT, and to use the CT to measure neutrino backgrounds below the limb and search for neutrinos through the Target of Opportunity searches as described in Section \ref{sec:Mission-Science-Goals}. On launch day of EUSO-SPB2, the high altitude air circulation pattern was well-established as demonstrated by the textbook circumnavigations (Figure~\ref{fig:Trajectory}) of the first mission~\cite{SuperBIT:2024a} of the 2023 W\=anaka campaign. That balloon, launched 27 days earlier, was in flight and performing well.

The EUSO-SPB2 first-night plan prioritized instrument turn-on and telescope commissioning to explore trigger rates and other metrics with first-light science operations as conditions permitted. The transition to EAS observations and neutrino searches was planned for subsequent nights. For the CT this meant dividing time between above the limb observations of HECRs at different telescope elevation angles together with a schedule of ToO follow-up of potential sources of high energy neutrinos that would require tilting in elevation angle and slewing in azimuth angle. Three collaboration shift centers at different timezones were standing by.

\subsection{Launch and Flight Duration}
 EUSO-SPB2 launched on 13 May$^{\text{th}}$, 00:02 UTC, a few days past the new moon, with 5 hours of moon-down time available the first night. The ascent was nominal, with the payload reaching the target float altitude of 33.5 km in about two hours. All systems, except for the FT HVPS and CT SiPM bias voltage, were powered before launch and throughout the flight. 

\begin{figure}[h!]\begin{centering}
	\includegraphics[width=0.65\textwidth]{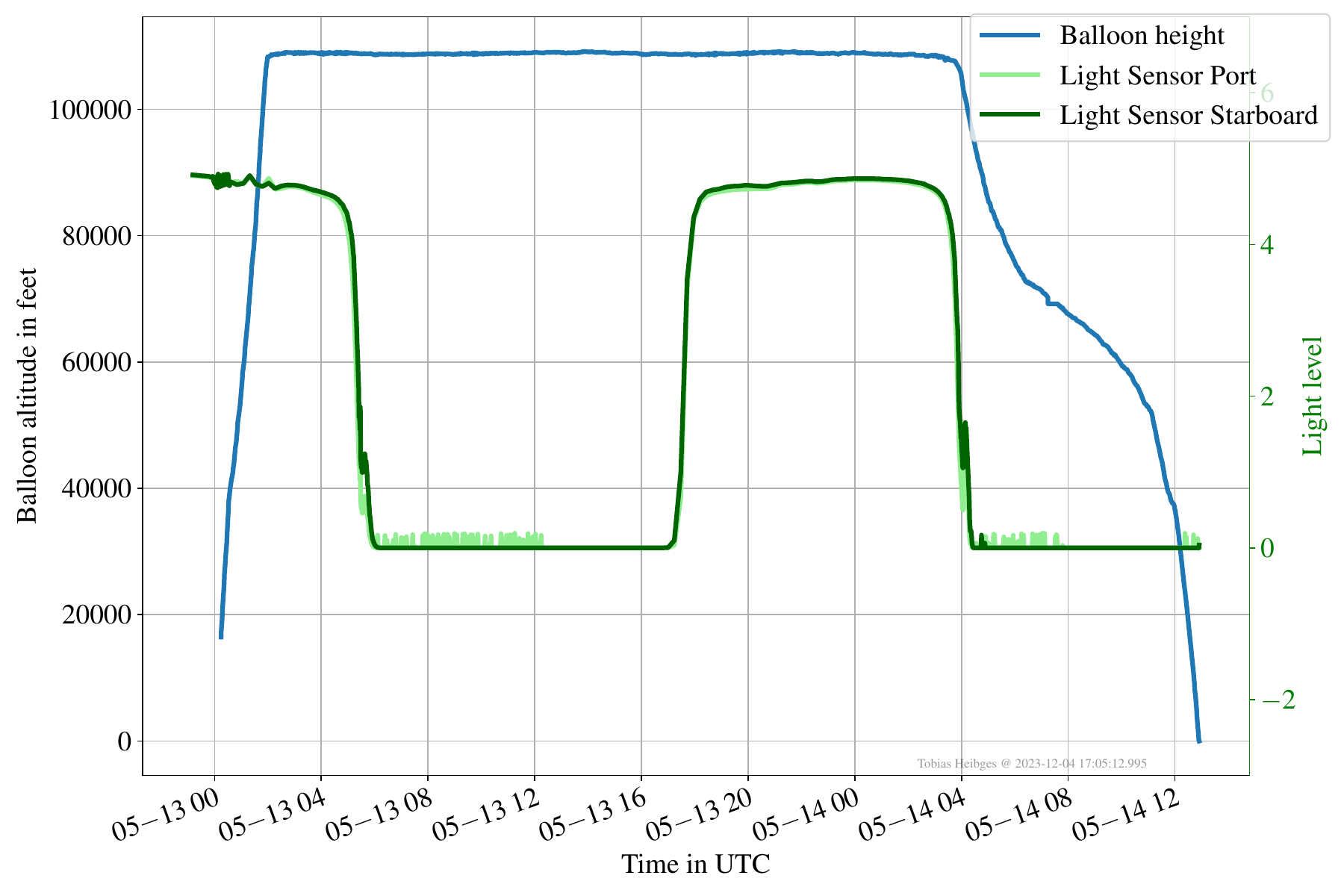}
	\caption{Balloon altitude from launch to termination. The light level output of two photocells mounted on the payload shows day and night (level = 0) periods.}
	\label{fig:altitude_in_flight}
\end{centering}
\end{figure}

Unfortunately, the flight lasted just 36 hours and 52 minutes, and ended with the payload loss in the ocean. The balloon altitude profile is plotted in Figure~\ref{fig:altitude_in_flight} with light sensor data superimposed to indicate night and day periods at the payload. The balloon maintained a super-pressure state of constant altitude during the night of the 13th. The science team was informed of a balloon leak the next morning. Unfortunately, by the start of the second night aloft (14 May), the balloon had already descended by 3~km. About 8 hours later the entire flight train splashed down. Releasing all 600 lbs of ballast failed to stop the descent and indicated that a leak was present in the upper part of the balloon.

\subsection{Operations and Data Collected}
Despite the short time available, the CT, FT, and IR camera systems were successfully commissioned and performance and science data were downloaded from each instrument. A NASA-supplied maritime Starlink unit, first integrated in W\=anaka, provided 100 times more bandwidth than the NASA TDRSS system for downloading EUSO-SPB2 data (Table \ref{tab:ddown}). 

\begin{table}[htp]
\centering
\begin{tabular}{| l l l|}
\hline
System& TDRSS & Starlink\\
\hline
CT  & 168 MB& 16,506 MB \\
FT  & 302 & 40,441\\
IR  & 61& 450 \\
\hline
Total & 531 & 57,397\\
\hline
\end{tabular}
\caption{EUSO-SPB2 data as downloaded from each system by telemetry link. }
\label{tab:ddown}
\end{table}

\subsubsection{Solar Power System} 
\begin{figure}[h!]\begin{centering}
	\includegraphics[width=0.65\textwidth]{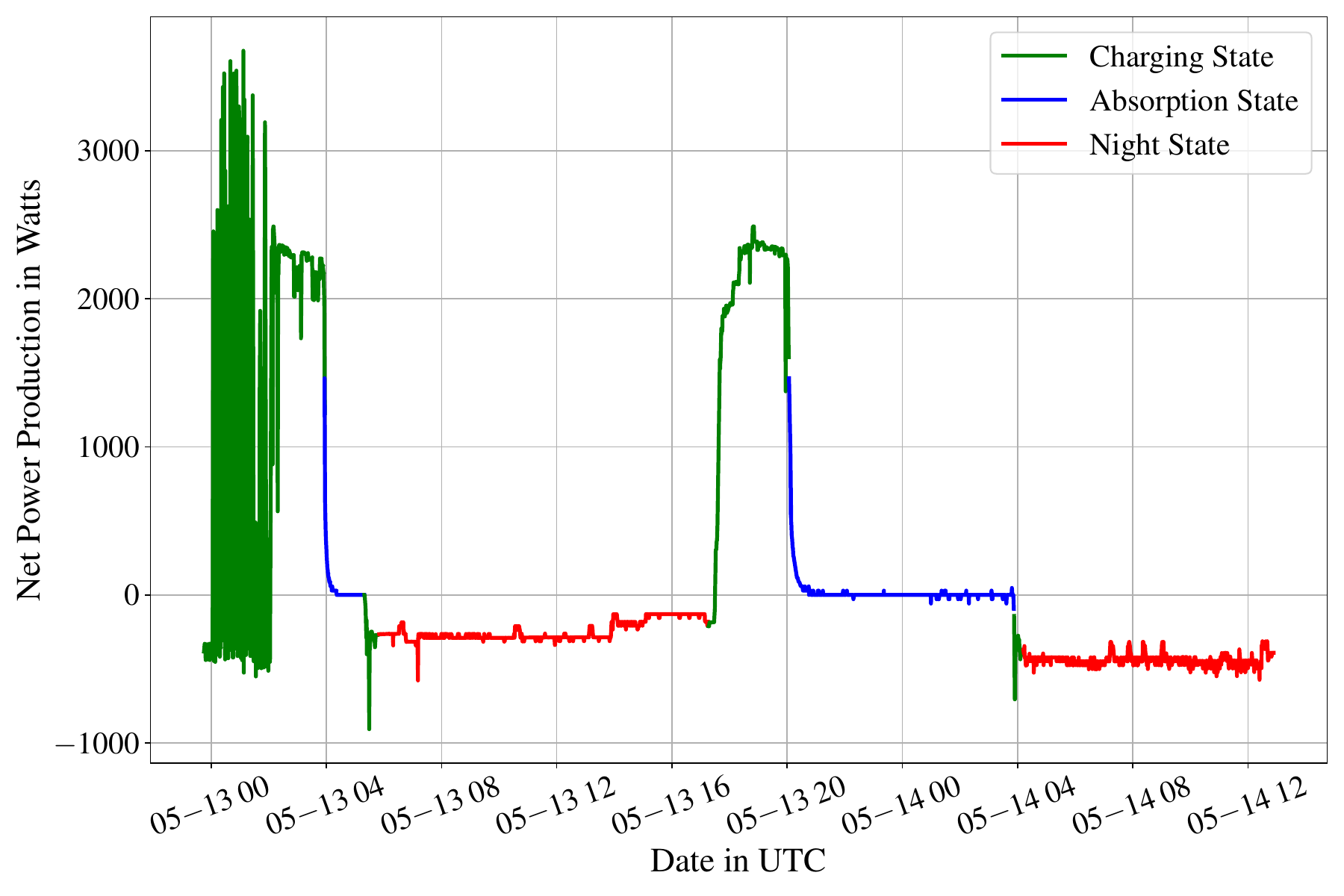}
	\caption{Net power of the solar power system. }
	\label{fig:solar_in_flight}
\end{centering}
\end{figure}

At float altitude, the solar power system generated about 2300 watts as the batteries were charged. In the cold temperatures and brighter sunlight of near space, the system produced about 50\% more power than its 1500 watt ground rating. During ascent power production of solar power system (Figure \ref{fig:solar_in_flight}) fluctuated while flight train slowly rotated and ``science" solar array swept past the Sun. This oscillation was largely removed after the NASA rotator was activated to enable daytime Sun tracking of the array.  

\subsubsection{Sample Temperature Data}
\begin{figure}[h!]
  \centering
  \includegraphics[width=0.60\textwidth]{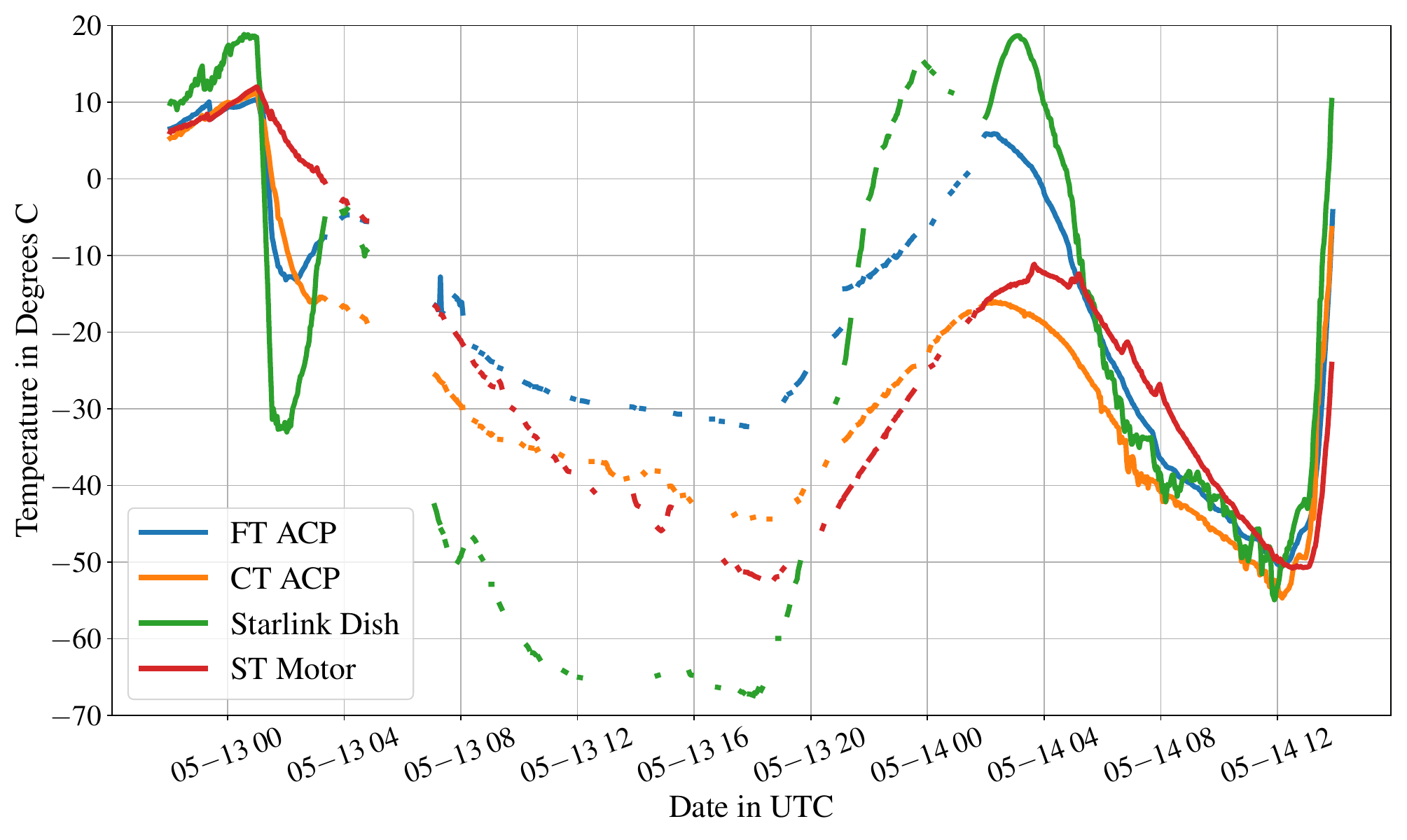} \\
  \includegraphics[width=0.60\textwidth]{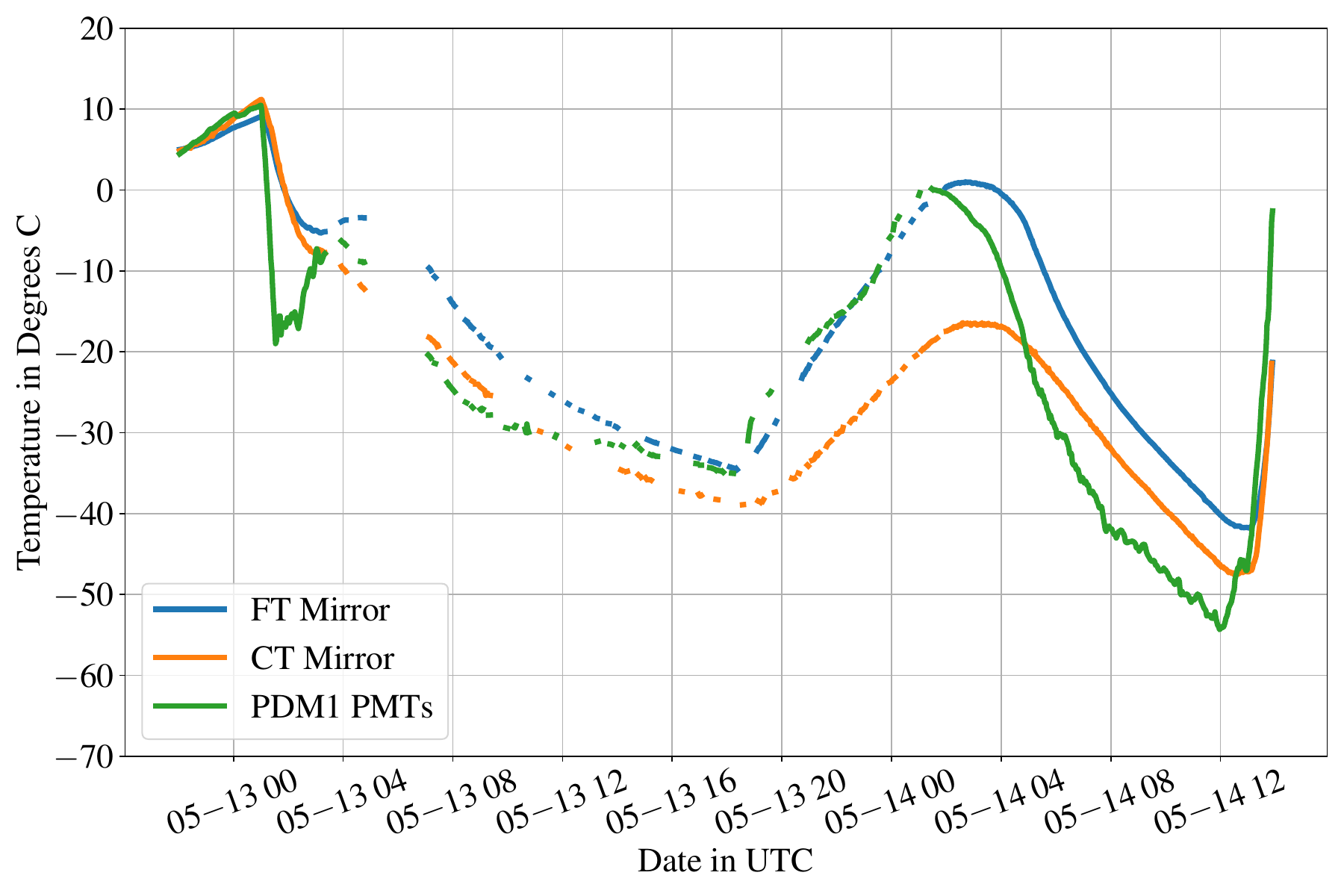} 
  \caption{Temperatures of selected components outside (top panel) and inside (lower panel) of the payload over the flight} 
  \label{fig:temps_in_flight}
\end{figure}

Temperatures outside the telescopes (Figure \ref{fig:temps_in_flight}) ranged from highs of slightly less than $+20\degree$ C to a recorded low of $-65\degree$ C on the Starlink antenna mount. The stepper motor in the tilting mechanism had a tight mechanical coupling to the gondola structure and dipped below -50\degree C. Inside the telescopes, temperatures of the FT and CT mirrors at float altitude ranged from $+10\degree$ C~to $-40\degree$ C (CT) and to $-35\degree$ C (FT). The average temperature of the CT SiPMs (not shown) ranged from +4\degree C to -6\degree C. The flight was too short to establish long-term temperature patterns.

\newpage
\subsubsection{IR Camera System}
\begin{figure}[ht]
    \centering
    \includegraphics[width=0.75\textwidth]{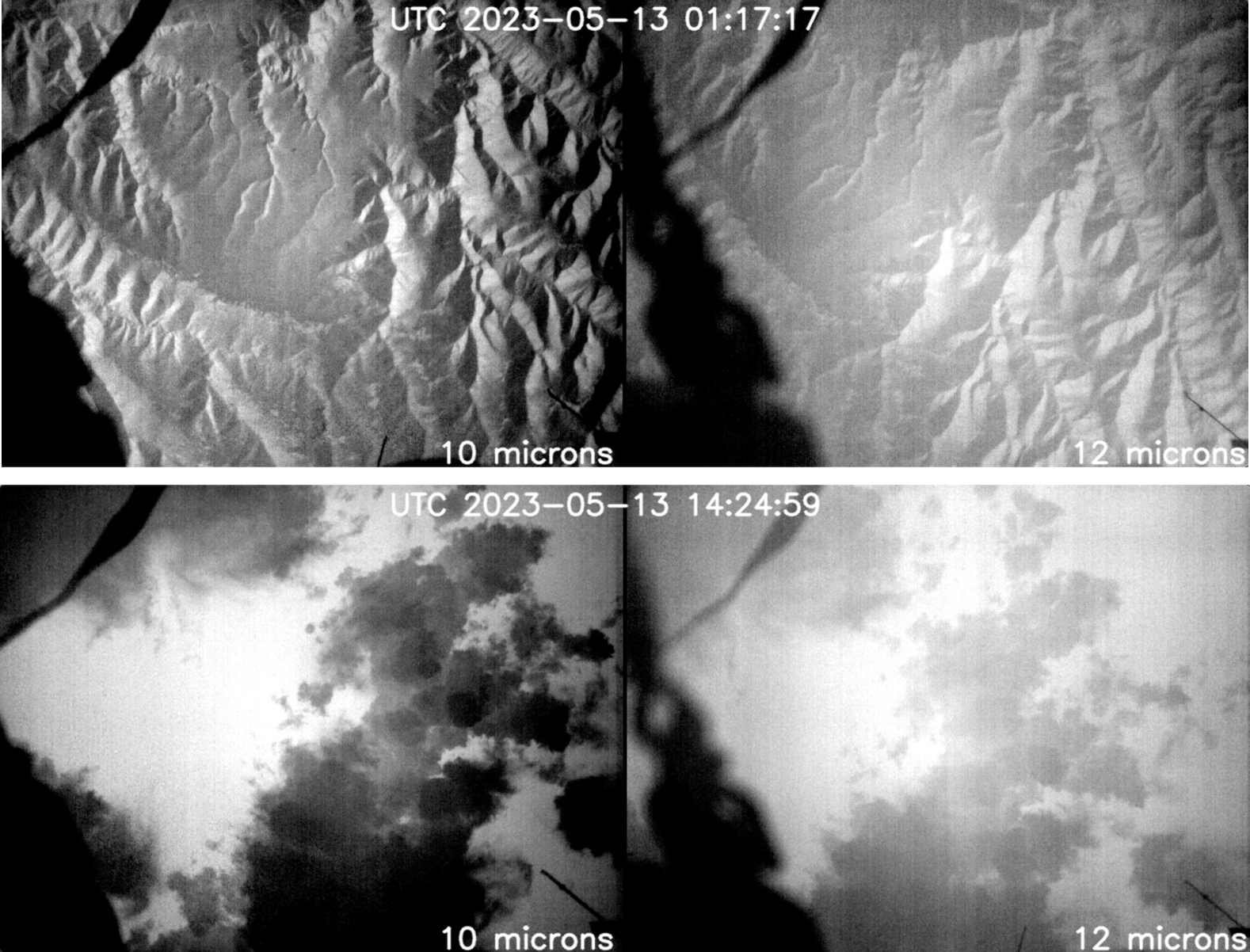}
    \caption{Uncalibrated sample images of mountains (top) during the ascent, 65 minutes after launch, and clouds as seen from 33.5 km (bottom) captured by the UCIRC2 IR camera system. Visible in the foreground is a portion of the EUSO-SPB2 gondola (lower left corners of each image) as well as cables and antennas hanging from the gondola.}
    \label{fig:SamplePic}
\end{figure}

\begin{figure}[ht]
    \centering
    \includegraphics[width=0.75\textwidth]{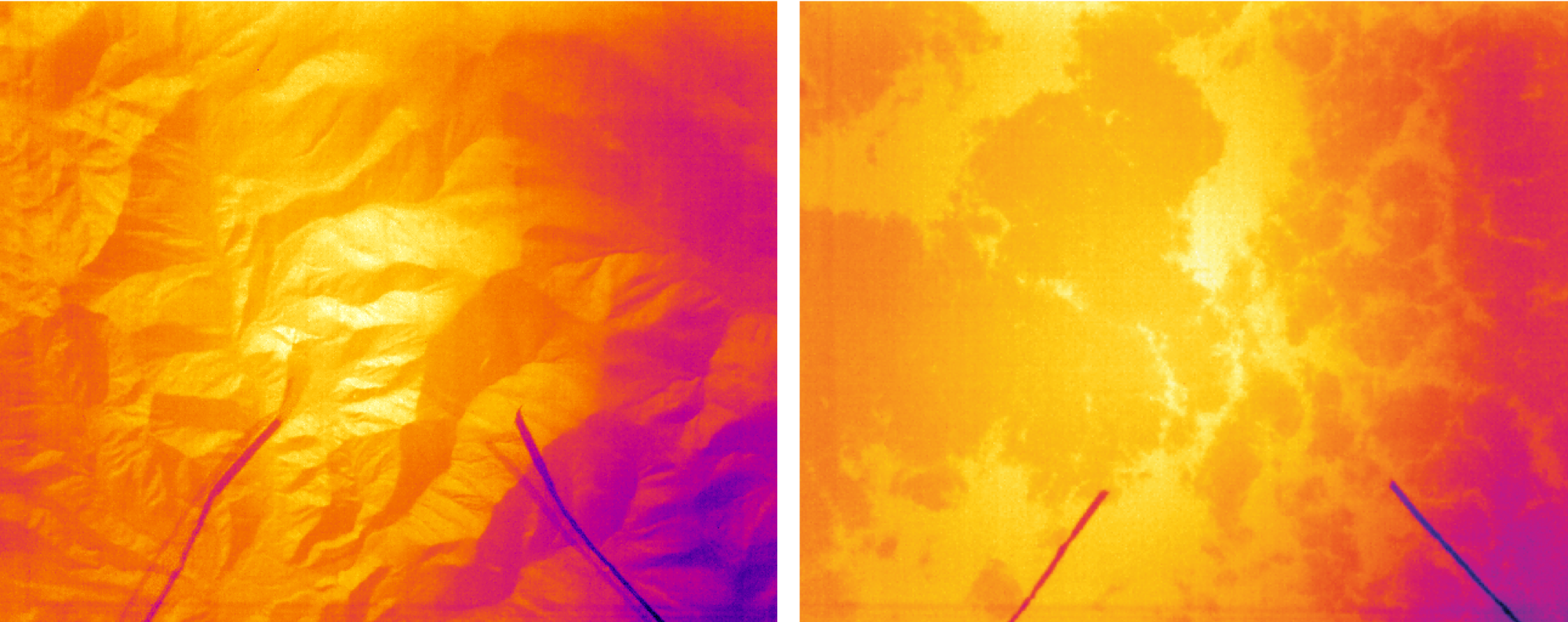}
    \caption{Sample images from the UCIRC2 IR camera system after calibration}
    \label{fig:UCIRC_Calibrated_Images}
\end{figure}

The IR Camera system recorded pictures every two minutes during the flight with its two cameras. 
Example camera images, two during ascent that captured mountains and two at float that captured cloud patterns, are shown in Figures \ref{fig:SamplePic} and \ref{fig:UCIRC_Calibrated_Images}.

\subsubsection{FT system}
The FT housekeeping monitoring worked as expected. During camera commissioning on the first night, about half of the EC modules in the camera reached their nominal voltage of 1013~V.  After launch, there were stability issues with the Starlink telemetry connection, which limited our download capacity and increased latency in the communication with the instrument. 
However, this was fixed on the following day and did not inhibit operational ability during that first night. Data collection continued under these conditions for the remainder of the dark period on 13 May$^{\text{th}}$, with plans to diagnose HVPS problems the following night, using full telemetry. 

\begin{figure}[h!]\begin{center}
	\includegraphics[width=0.8\textwidth]{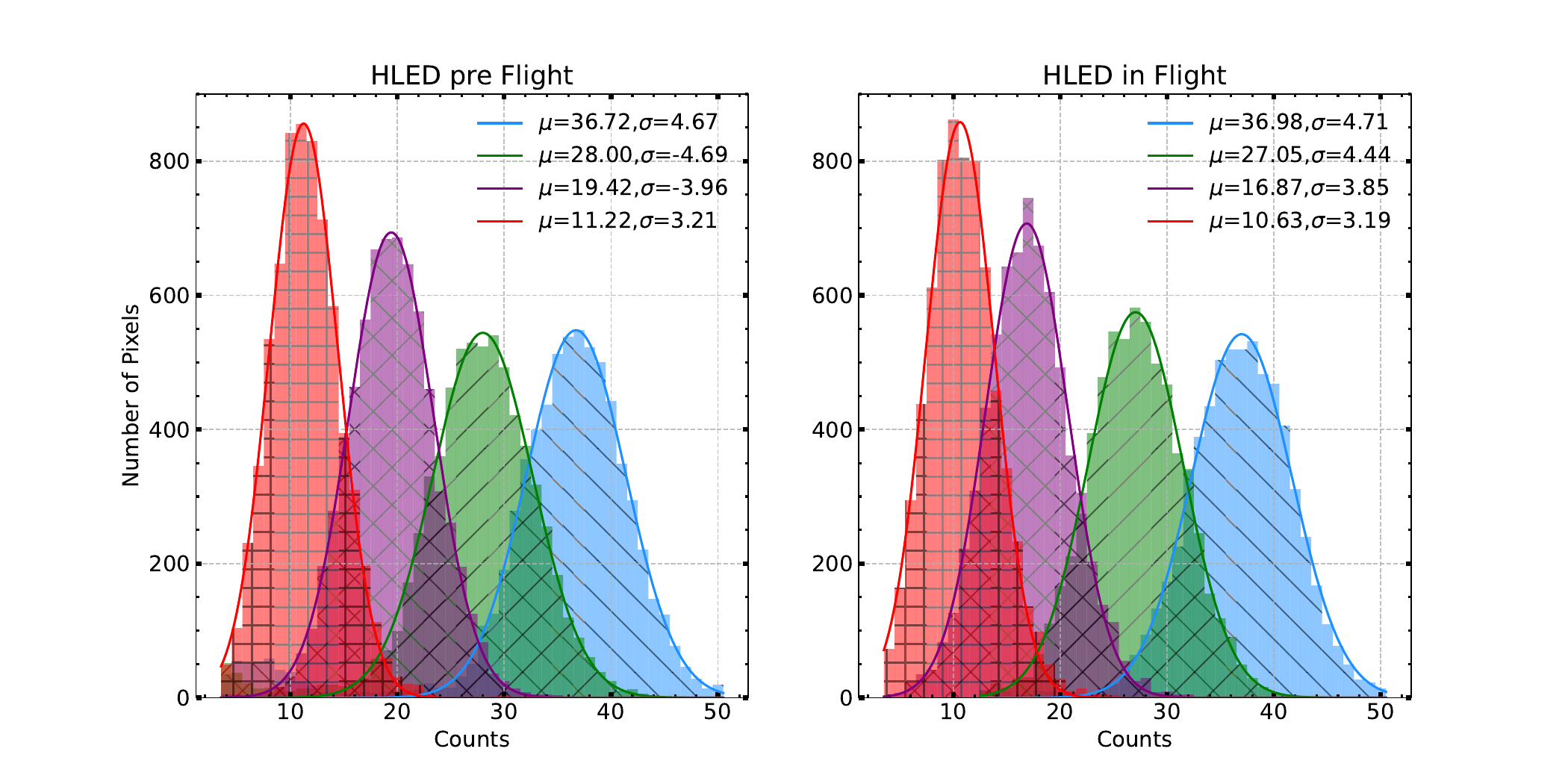}
	\caption{Number of pixels above threshold illuminated by the HLED as a function of photoelectron counts recorded preflight (left) and inflight (right). Different flashes, of different intensities, are shown in different colors. Solid lines represent three parameter Gaussian fits \citep{Filippatos:2023P4}.}
	\label{figs:FT_hled_flight}
\end{center}\end{figure}

A modified power-on sequence was implemented on the second night that ramped up the HV over a longer period of about 2.5 minutes. After this change, all 27 EC modules turned on at the intended potential difference and at nominal efficiency. After performing a threshold scan, the instrument operated for the remainder of the flight. In total 99,682 triggered events were recorded and downloaded. As a demonstration of system performance, measurements of the health LED collected on the ground and at float are compared in Figure \ref{figs:FT_hled_flight}.
  
\begin{figure}[h!]\begin{center}
	\includegraphics[width=0.9\textwidth]{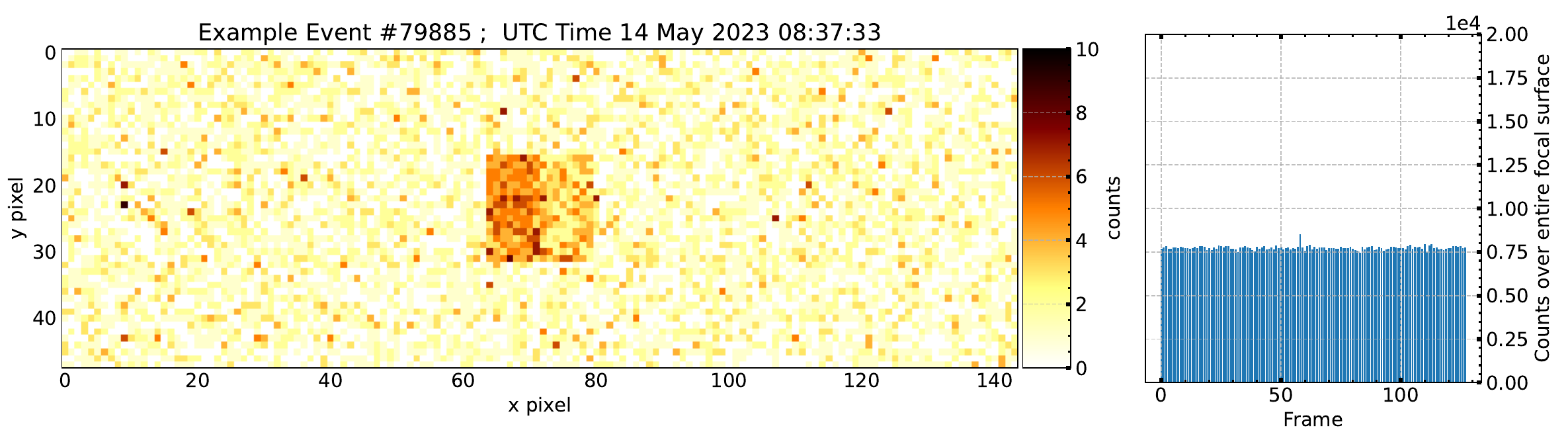}
	\caption{{A example event trigger recorded during the second night of the flight. The pattern of higher signal follows the layout of one 8x8 pixel MAPMT within the focal surface. Time profile of the event is shown in the right panels, with the brightest frame shown in the left panels (see text).}}
	\label{figs:example-ft-2}
\end{center}\end{figure}

Most of the FT events recorded during the short flight were contained in single 1 $\mu$s frames. 
These events may be explained by a charged particle interacting directly with the focal surface, causing an excess of hits to be recorded. An example of a particularly bright event is shown in Figure \ref{figs:example-ft-2}, along with an event of typical brightness. The signal in this example is spread over multiple MAPMTs and two PDMs; however, many events are localized to a singular EC.

Within the FT events downloaded, searches for track-like EAS signatures have used neural network-based approach and a more ''traditional" track-finding approach under the JEM-EUSO Offline framework \cite{JEM-EUSO:23fyg}. 
No EAS candidates were found, which is consistent with the expectation of 0 to 2 events over the limited integrated exposure for the curtailed flight. An observation time of slightly less than three hours was identified by selecting the times between observed HLED events. The balloon was descending during part of the period, leading to a decrease in peak sensitivity from 2.4 EeV at 33 km to 0.8 EeV and an increase in expected detection rate from 0.2 to 0.8 events/hr.  Details of this analysis, including examples of non-EAS event triggers, can be found in ~\cite{EUSO:SPB2FT:2024a}. 

\begin{figure}[h!]\begin{centering}
	\includegraphics[width=0.8\textwidth]{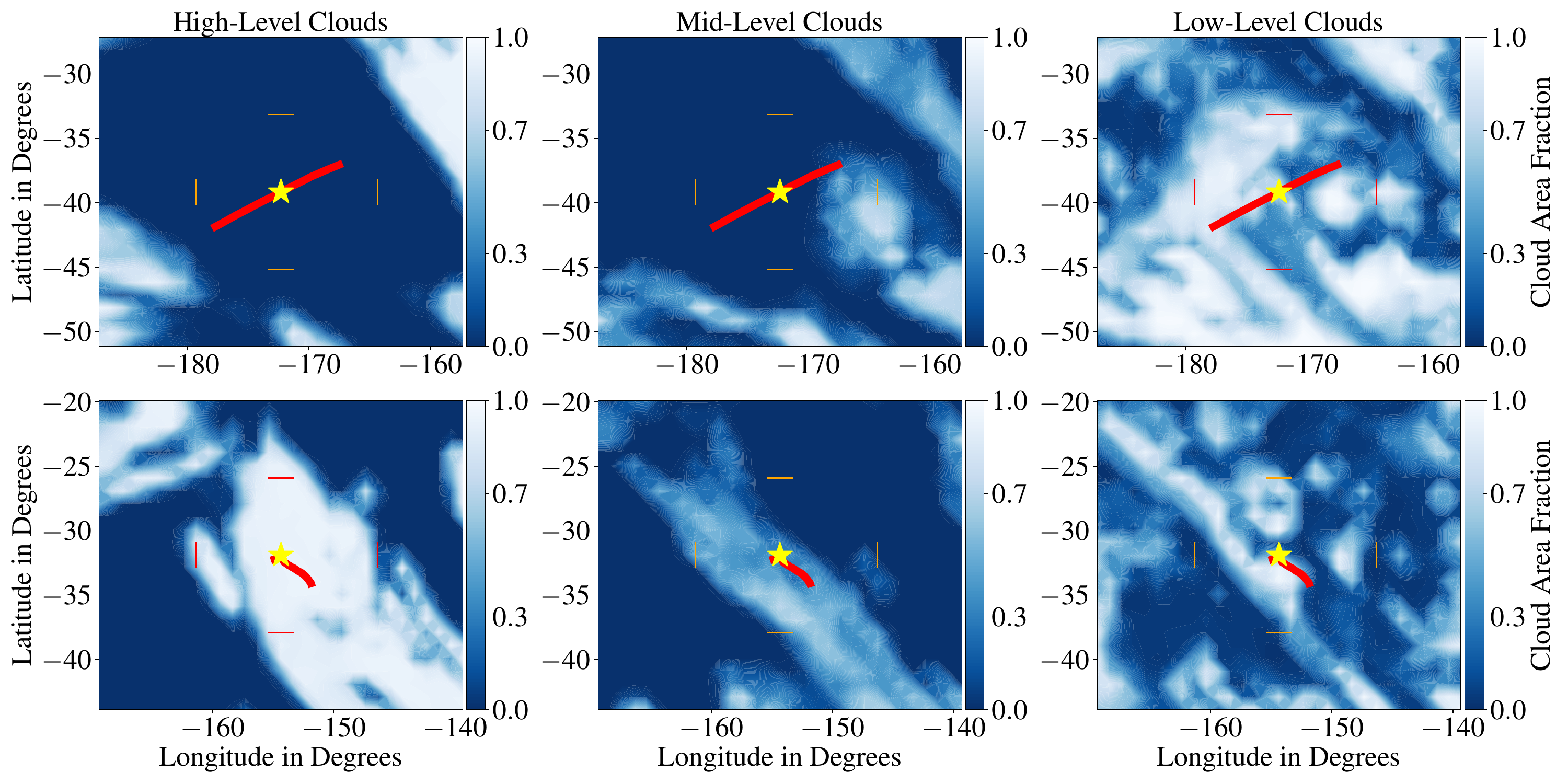}
	\caption{Cloud coverage maps, extracted from NASA MERRA-2 data, for the two nights of the flight. The three upper panels correspond to the night of May 13 at 12 UTC. The three lower panels correspond to May 14 at 6 UTC. The yellow stars indicate the position of the balloon at these times. Maps for three altitude ranges (see text) shown from left to right (see text). The balloon trajectory for the night is indicated by the thick red line. The thin vertical and horizontal marks indicate the 650 km distance to the Earth's limb as seen from the nominal 33.5 km balloon height.}
	\label{fig:Flight_Clouds_NASA}
\end{centering}
\end{figure}

As a follow-up to this study, large area mapping of cloud coverage was retrieved post-flight from NASA's Modern-Era Retrospective analysis for Research and Applications, Version 2 (MERRA-2) \cite{TheModernEraRetrospectiveAnalysisforResearchandApplicationsVersion2MERRA2} combined dataset. Cloud coverage maps created from MERRA-2 data\cite{MERRA-2-File-Used} for the two nights of the flight are shown in Figure \ref{fig:Flight_Clouds_NASA}. The latitude and longitude ranges displayed extend roughly two horizon distances, or about 1300 km, from the balloon position. Maps are shown for three different altitude zones. At high-level (5-14 km) and mid-level (2-7 km) altitudes, the sky was largely cloud-free during the first night. High-level clouds and mid-level clouds are present in the second night. Obscuration due to these clouds likely contributed to the null result of the EAS search because the data used was collected on the second night. Both nights had low-level (0-2 km) clouds.

\newpage
\subsubsection{CT system}

\begin{figure}
\begin{minipage}[c]{0.48\linewidth}
\includegraphics[width=\linewidth]{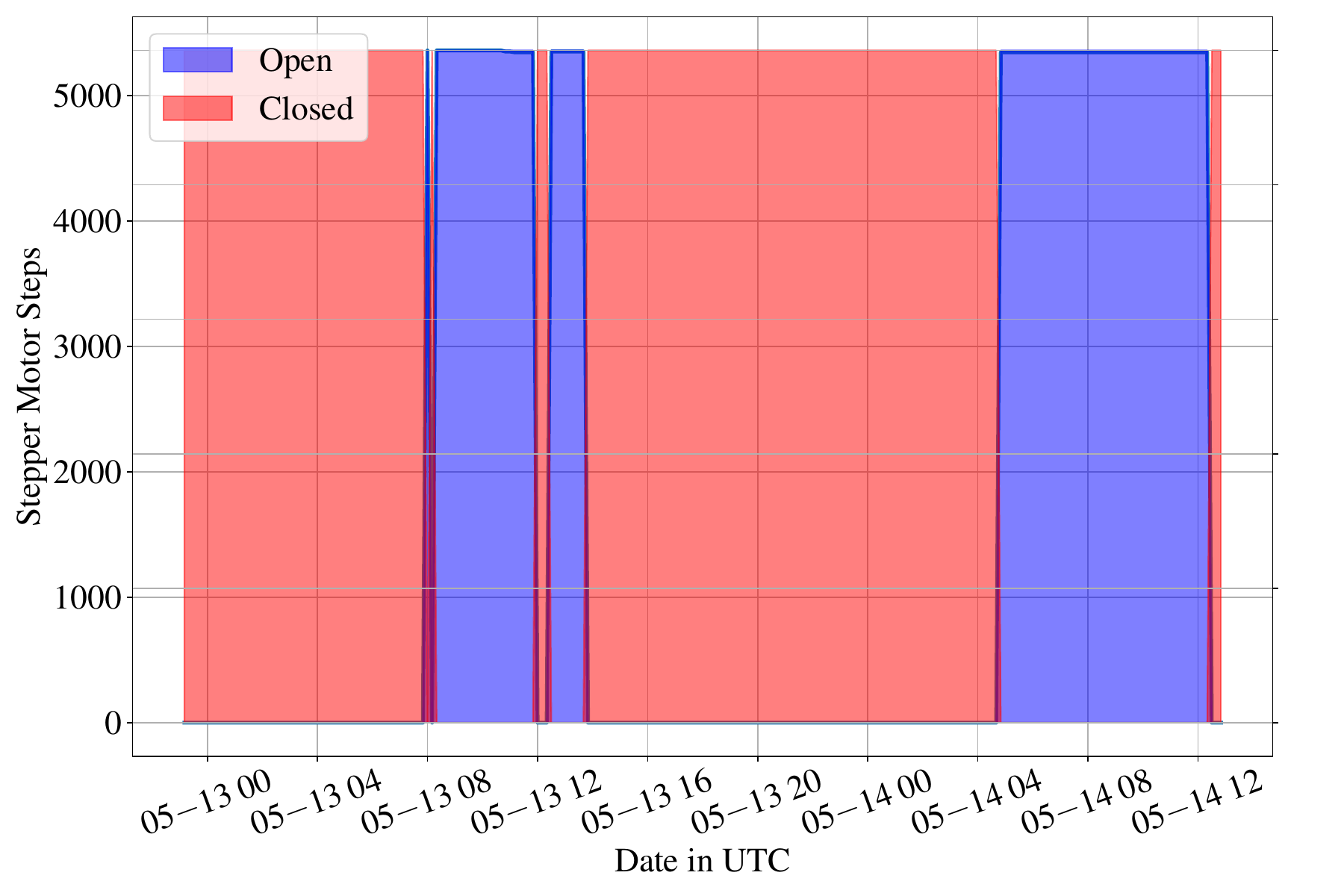}
\caption{CT shutter status over the mission.}
	\label{figs:shutter_opps}
\end{minipage}
\hfill
\begin{minipage}[c]{0.48\linewidth}
 	\includegraphics[width=\linewidth]{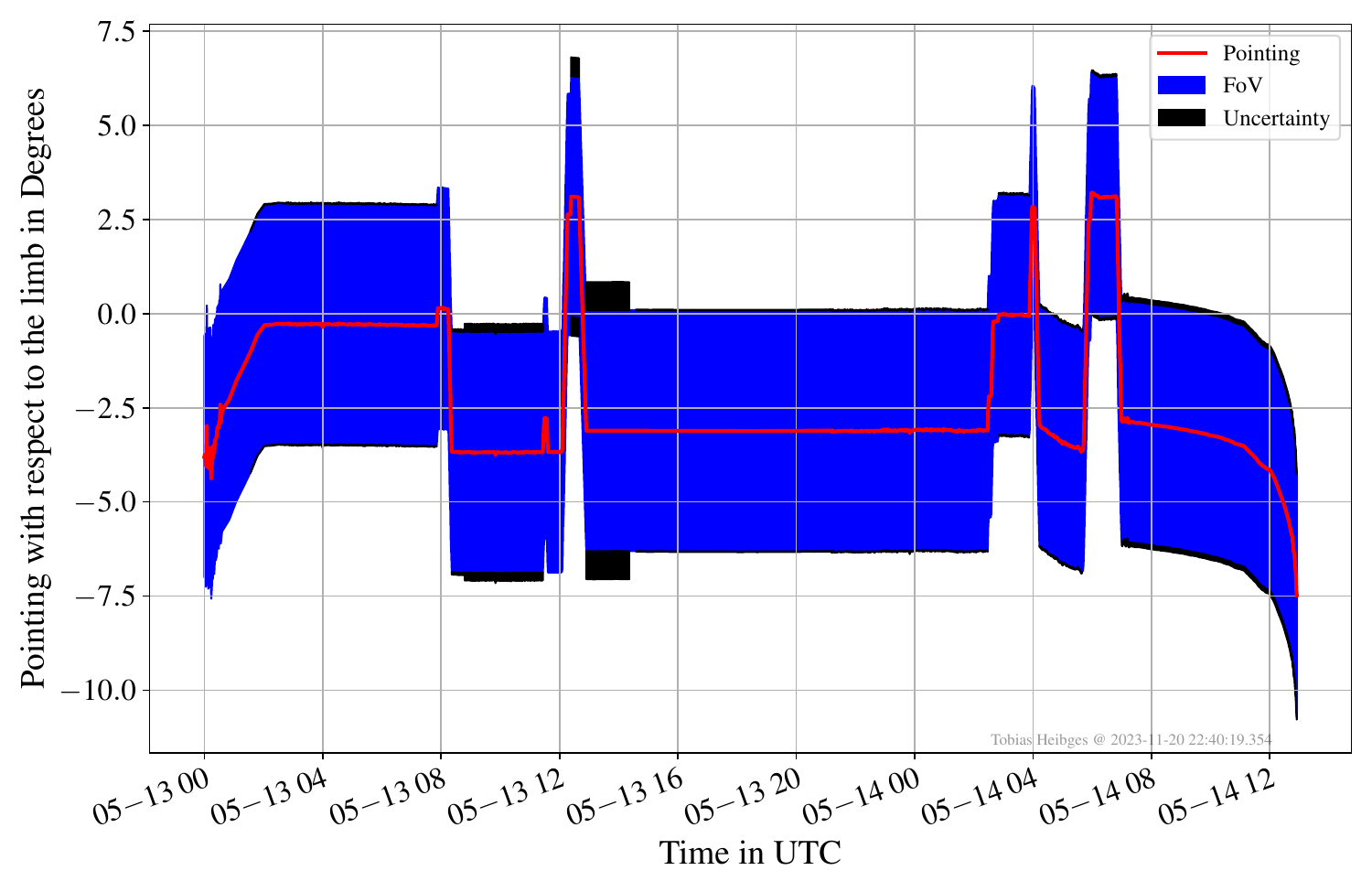}
    \caption{CT vertical FoV range over the flight. The red curve indicates the tilt angle as defined in the text.}
	\label{figs:tilt_limb}
\end{minipage}
\end{figure}

The CT system was commissioned on the first night of flight. The camera system was first turned on with the shutters closed and HLED data were collected. The shutters were opened for data collection at 12 UTC after a brief open / closed test cycle. The shutter status on the mission is plotted in Figure \ref{figs:shutter_opps}. Telescope operations on the first night included an exploration of trigger rates as a function of SiPM bias voltage and bi-focal threshold settings and pointing direction, including a full 360\degree~sweep in azimuth. The first data with the telescope pointed above the limb were also collected. The instrument was also operated pointing north after the moon (51\% illumination)  rose to the east, illustrating the advantage of enclosing the entire Cherenkov telescope except the entrance pupil.

\begin{figure}[h!]\begin{centering}
	\includegraphics[width=0.50\textwidth]{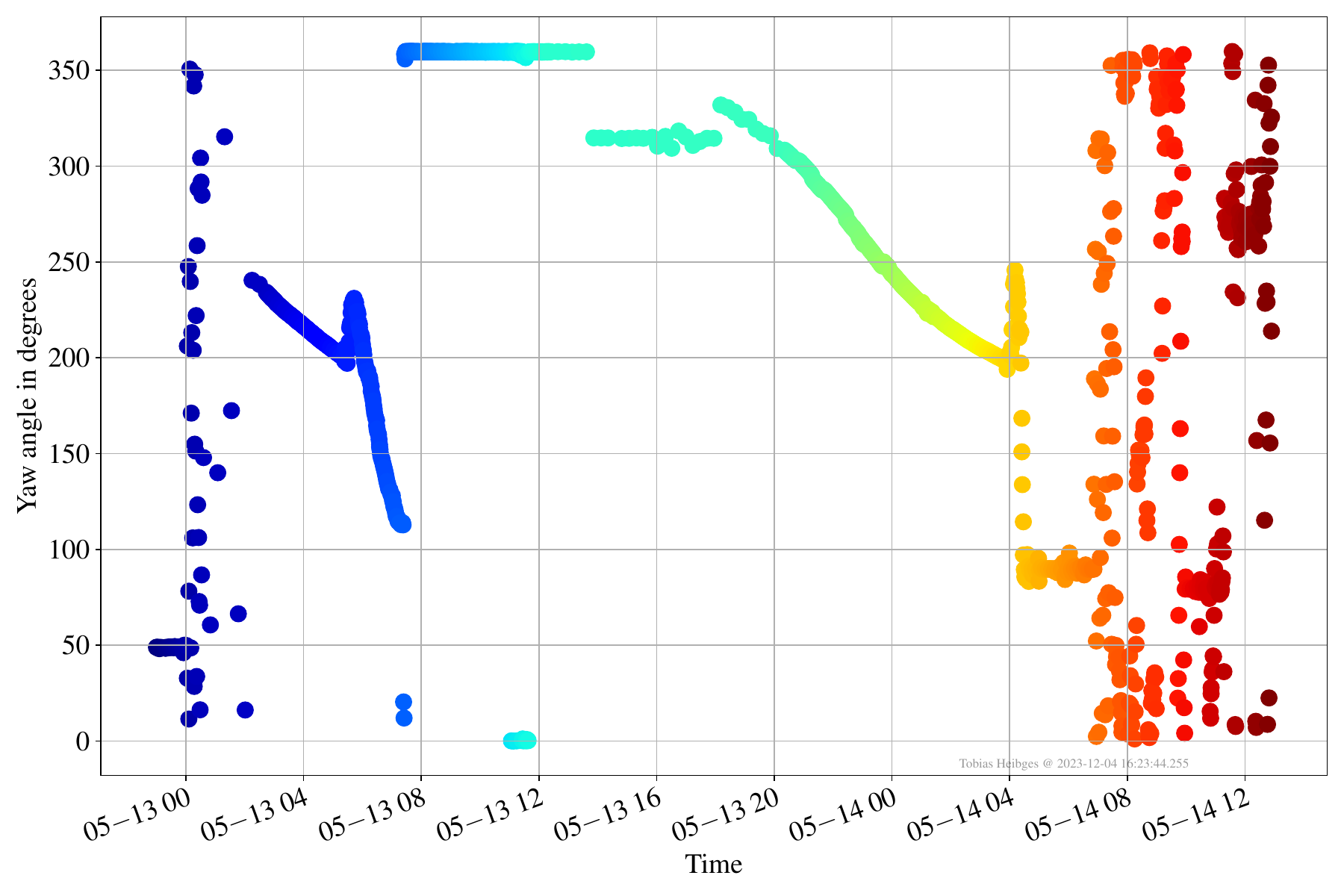}
	\caption{The azimuth pointing direction of the CT displayed over the flight.}
	\label{fig:azimuth_in_flight}
\end{centering}
\end{figure}

The tilt angle of the CT optical axis and the CT vertical field of view, both relative to Earth's limb during flight, are shown in Figure \ref{figs:tilt_limb}. The latter includes the change in elevation of the balloon as it descended on the second night. The azimuth-pointing directions during the mission are illustrated in Figure \ref{fig:azimuth_in_flight}.

\begin{figure}
\begin{minipage}[c]{0.59\linewidth}
\includegraphics[width=\linewidth]{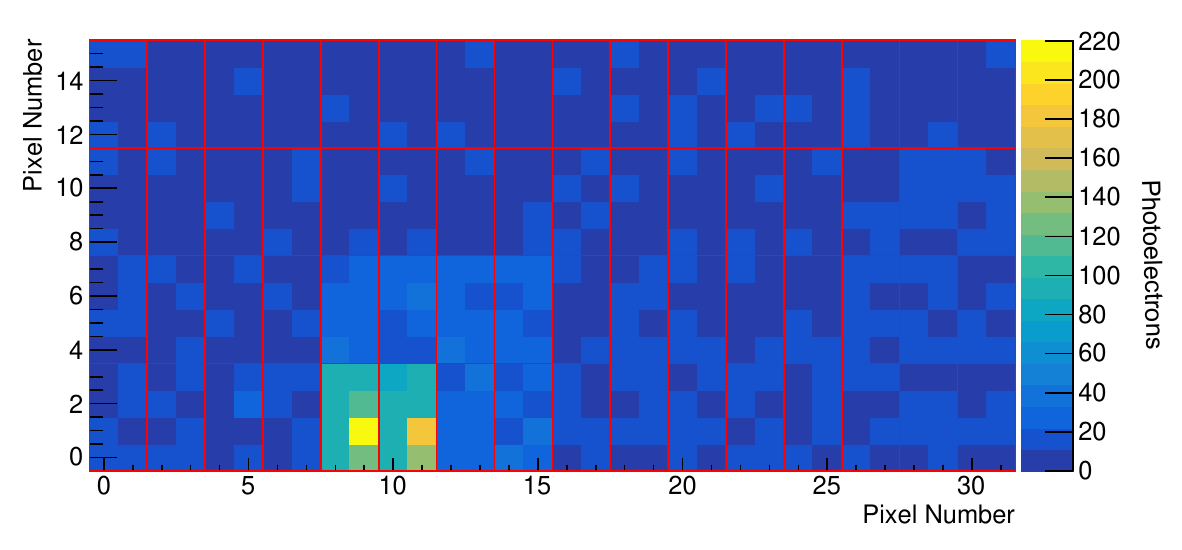}
\end{minipage}
\hfill
\begin{minipage}[c]{0.48\linewidth}
 	\includegraphics[width=0.66\linewidth]{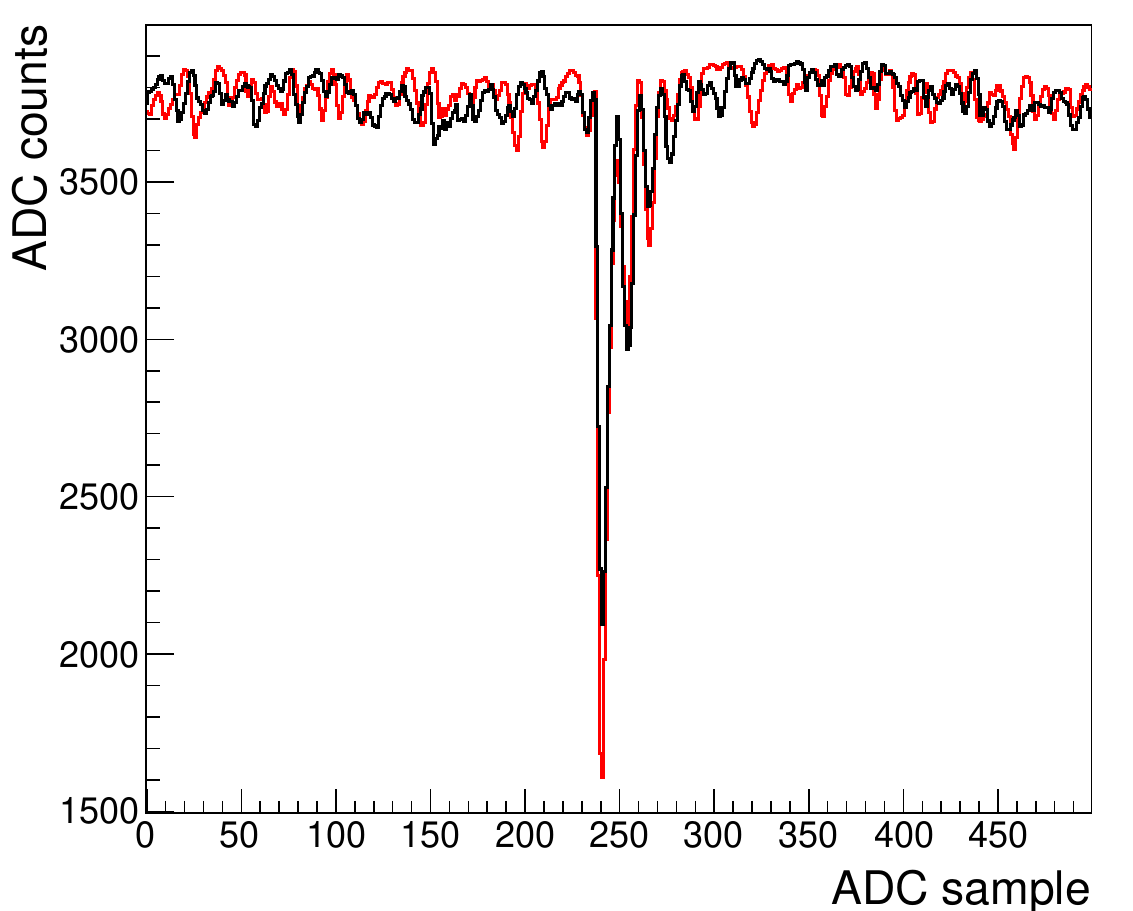}
\end{minipage}
\caption{Example of an EAS candidate observed Cherenkov Telescope on May 14 2023.  Left Panel: Camera view with photoelectron scale. Right Panel: Time traces of the two brightest pixels of the bifocal image. The bottom of the camera image corresponds to the top of the field of view. The two brightest pixels illuminated by this event were pointing 5.7\degree~above Earth's limb when this event was recorded.}
	\label{fig:Flight_CR_Candidates}
\end{figure}

On the second night, approximately 45 minutes of data with the telescope optical axis pointing 3\degree~above the limb were collected to search for HECRs which had been predicted for a sub-orbital mission. 2893 bifocal triggers were recorded. A search of these events identified about 10 HECR candidates \cite{EUSO:Gazda2023a}. An example is shown in Figure \ref{fig:Flight_CR_Candidates}. As was the case for on-axis HECRs observed in the field tests, the bi-focal time traces are correlated in time and well above background. The majority of these events are located near the top of the field of view of the CT where the smaller atmospheric slant depth yields and lower atmospheric attenuation yields a lower energy threshold. Publication of a detailed analysis and simulation of this data set is in preparation.

\begin{figure}[h!]\begin{centering}
	\includegraphics[width=0.8\textwidth]{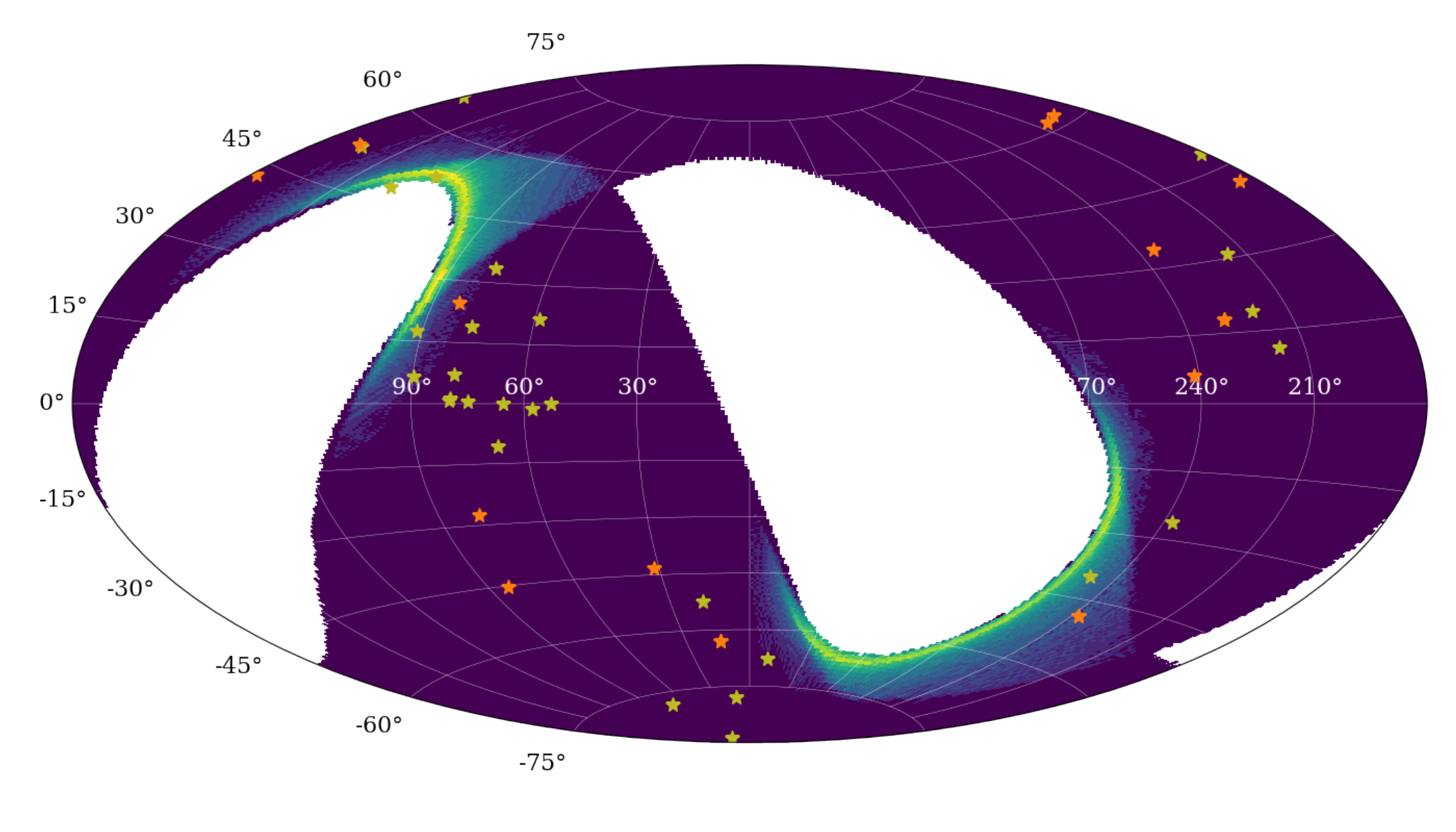}
	\caption{Skymap in galactic coordinates of ToO sources within the 6.4\degree~band below the limb as seen from the balloon for the observable period on the night of May 14th 2023. Stars indicate the positions of the ToO candidate sources. Yellow are transient sources, Orange are steady-state sources. The color of the region indicates geometrical acceptance. The brighter bands are 10 times higher than the rest of the region.}
	\label{fig:ToO_Sky_Map}
\end{centering}
\end{figure}

A set of ToO source candidates and a ToO observing schedule had been prepared for the second night of flight and subsequent nights. The source candidates were drawn from a catalog that was populated daily with selected alerts from the General Coordinates Network, the Transient Name Server, and the Astronomer's Telegram. A custom scheduler\cite{ToO:Scheduler:2024a} then prioritized the different candidate sources and produced a schedule of UTC times and pointing directions. These pointing directions corresponded to the optimal azimuth and elevation angle, as measured at the balloon, to catch the source as it crossed the CT FoV just below the horizon. The balloon position was obtained from NASA-provided trajectory extrapolations that were updated every few hours. The candidate sources that would rise or set during the dark period of the second night are mapped in galactic coordinates in Figure \ref{fig:ToO_Sky_Map}. From this list, the ToO scheduler had selected 5 sources.

Unfortunately, the balloon descended too quickly to perform a ToO pointing operation. By the time other CT operations had been completed, the balloon was descending through the tropopause, and rotating faster than the azimuth gondola rotator was designed to handle. The balloon made 7 continuous rotations during the descent. A demonstrative test of the ToO method using brief coincidence overlaps with ToO candidate sources is in progress.
\newpage

%% file: 99_Conclusions.tex
\section{Conclusions and Outlook}
\label{sec:Conclusions}
The EUSO-SPB2 payload was delivered on time and within the allocated mass budget for a NASA super-pressure balloon flight from W\=anaka NZ. The payload preparation included field tests in the Utah desert of both telescopes. 

The science objectives included the first observation of UHECRs via fluorescence from near space, the first observation of cosmic rays with a Cherenkov telescope from near space, measurement of night-sky backgrounds above and below the limb with a Cherenkov telescope from near space, and tests of pioneering methods to search for neutrinos using the Earth-skimming technique. These included the study of backgrounds for neutrino detection, the search for diffuse neutrinos below Earth's limb, and searches for tau neutrinos using the ToO methods.

A flight of up to 50 days was envisioned. Despite the setback of an aborted flight that ended in the Pacific Ocean after 37 hours, all instruments, the solar power system, and the Starlink telemetry unit were successfully commissioned. The 56 GB of data downloaded included about 100,000 FT triggers and more than 32,000 bifocal CT triggers. A search of the FT data for UHECRs came up empty, consistent with the expectation of zero to two events for a flight of this duration. About 10 HECR event candidates have been identified from 45 minutes of above-limb CT observations. This observation demonstrated, in situ, that the CT was working and supported the expectation that when tilted below Earth's limb to look for high energy astrophysical tau neutrinos it would have sensitivity to EASs, which are the predicted detection signature. Analysis and publication of the below-limb data to set limits on the diffuse neutrino flux and measure background light levels is in progress. There was insufficient time to demonstrate a ToO pointing operation, although schedules were prepared and standing by. A paper demonstrating the method through brief coincidental observations of lower priority ToO sources that crossed the CT FoV below the limb is in progress.

The EUSO-SPB2 design and flight also motivated the development of the POEMMA Balloon with Radio (PBR) follow-up mission. PBR will include a hybrid Cherenkov/fluorescence telescope and a radio detector to pioneer simultaneous optical and radio EAS measurements and neutrino searches from near-space. PBR is currently in preparation for a launch from W\=anaka NZ in 2027.